\documentclass{aa}  

\usepackage{graphicx}
\usepackage{txfonts}
\usepackage[colorlinks=true,linkcolor=blue,citecolor=blue]{hyperref}
\begin{document}

   \title{Magnetic field amplification in massive primordial halos}

   \subtitle{Influence of Lyman-Werner radiation}

   \author{V. B. Díaz\inst{1}\fnmsep\thanks{vdiaz@hs.uni-hamburg.de}
          \and
          D. R. G. Schleicher\inst{2}
          \and
          M. A. Latif\inst{3}
          \and
          P. Grete\inst{1}
          \and
          R. Banerjee\inst{1}
          }

   \institute{Hamburger Sternwarte, Universität Hamburg, Gojenbergsweg 112, D-21029 Hamburg, Germany
         \and
             Departamento de Astronomía, Facultad de Ciencias Físicas y Matemáticas Universidad de Concepción, Av. Esteban Iturra s/n Barrio Universitario, Casilla 160-C, Chile
         \and
             Physics Department, College of Science, United Arab Emirates University, PO Box 15551, Al-Ain, UAE
             }

   \date{Received -; accepted -}

 
  \abstract
   {The potential importance of magnetic fields during structure formation and gravitational collapse in the early Universe has been shown in several studies. In particular, magnetic field amplification by the small-scale dynamo plays an important role in addition to the pure amplification expected from gravitational collapse.}
   {In this paper, we study the small-scale dynamo for halos of $\gtrsim10^7$~M$_\odot$ collapsing at $z\gtrsim12$, under different ambient conditions due to the strength of the Lyman-Werner background. Additionally, we estimate the approximate saturation level by varying the initial magnetic field strength.}
   {We performed cosmological magnetohydrodynamical simulations for three distinct halos of $\sim10^7$~M$_{\sun}$ at $z\geq13$ by varying the Jeans resolution from $32-256$ cells and employed Lyman Werner background flux of strengths $10^2-10^5$ in units of $J_{21}$, where $J_{21}=10^{-21}$~erg$/$cm$^2/$sr$/$s$/$Hz. To follow the chemical and thermal evolution of the gas we made use of the KROME package.}
   {In addition to the compression by collapse, we find magnetic field amplification via the dynamo both in the regimes of atomic and molecular hydrogen cooling. Moreover, we find a lower saturation level in the molecular hydrogen cooling regime. This behaviour can be understood due to the generally reduced radial infall velocities and vorticities in this regime, as well as the higher Mach numbers of the gas, which give rise to a smaller saturation ratio.}
   {Our results overall suggest that the dynamo operates over a large range of conditions in the collapsing gas.}


   \maketitle
%

\section{Introduction}
It is known that supermassive black holes (SMBHs) are present in the nuclei of most galaxies and through high redshift surveys their existence with masses of about $10^9$~M$_\odot$ was revealed at $z\geq6$ \citep{morlock,banados,domi2018}. This implies that they existed when the Universe was less than 800 Myr old and the formation and nature of their seeds are one of the mysteries of the Universe that remains to be solved. Now, with the upcoming observations thanks to the launch of new telescopes such as the James Webb Space Telescope (JWST) the number of these objects will increase, as it has been shown that it could detect seeds of the first quasars at $z\lesssim20$ \citep{Whalen2020}. In fact, recent observations of the JWST have already allowed the detection of massive and bright galaxies at very high redshift ($z\sim10$) which has also now been confirmed spectroscopically \citep{Curtis2023,Robertson2023,Arrabal2023}. In the coming decade, with surveys by Euclid and Nancy Grace Roman Space Telescope (RST) it will be possible to find even more of those objects at $z\lesssim6-8$, and even at higher redshift thanks to gravitational lensing \citep{Vikaeus2023}. Furthermore, \cite{Latif2023Radio} showed that the first quasars could also be detected by the Square Kilometer Array (SKA) up to $z\sim16$ and by the Next-Generation Very Large Array (ngVLA) at $z\sim14$. This leads to the need to further update the theory behind the formation of these massive objects with respect to these upcoming observations. The models that have been proposed to potentially explain the formation of SMBHs include the merging and accretion of Pop III remnants \citep{abel,yoshida,latif1}, the collapse of dense stellar clusters \citep{omukai1,devicchi1,tjarda,reinoso} and the direct collapse of protogalactic gas clouds in which the primordial gas in a massive halo collapses directly into a central and massive object $(\sim 10^5$~M$_\odot)$ without fragmentation \citep{brom,begelman,latif2,shlosman}. This last model seems to be the most promising scenario as it provides the most massive black hole seeds, which then can grow via moderate accretion rates to form SMBHs \citep{latif3}. The most likely cradles to assemble these direct collapse black holes (DCBHs) are the so-called atomic cooling halos which are metal-free very massive halos with virial temperature $\geq10^4$~K at $z\sim15-10$.

The main coolant in the early Universe is the molecular hydrogen (H$_2$), which efficiently cools down the gas to a few hundred K possibly triggering fragmentation and therefore impeding the atomic-cooling halos to collapse into a DCBH. It is therefore important to suppress the H$_2$ cooling process. In this respect, the presence of an intense UV flux produced by Pop~III or Pop~II stars is mandatory to inhibit the H$_2$ formation through direct dissociation of this molecule or by H$^-$ photodetachment \citep{domi2010}, which can be achieved if the halo is close to a star-forming galaxy \citep{Dijkstra1}.

The required UV background field to maintain the gas in an atomic state, i.e., the critical strength of the radiation flux $J_{21}^{crit}$, where $J_{21} = 1$ corresponds to the specific intensity just below the Lyman limit (13.6 eV) in units of $10^{-21}$~erg~cm$^{-2}$~sr$^{-1}$~s$^{-1}$~Hz$^{-1}$, has been obtained performing one-zone and three-dimensional (3D) simulations \citep{omukai1,shang,latif4}. The critical value depends on the chemical model and the spectral shape of the radiation field. This radiation field is commonly described as an idealized spectrum with blackbody radiation temperatures of T$_{rad} = 10^{5}$~K and T$_{rad} = 10^{4}$~K for Pop III and Pop II, respectively. The J$_{21}^{crit}$ parameter has a direct impact on the estimation of the DCBH number density ($n_{DCBH}$), affecting the feasibility of this scenario, so that it is of importance to estimate the value of J$_{21}^{crit}$ with high precision \citep{Dijkstra2,yue,Habouzit2016}. In this context, \citet{sugimura} estimated the value of J$_{21}^{crit}$ using a one-zone model including more realistic spectra, finding a critical value of J$_{21}^{crit} = 1000 - 1400$. Even more, \citet{latif5} performing 3D simulations and using a realistic spectra (T$_{rad} = 2 \times 10^{4} - 10^{5}$~K) found a higher value for J$_{21}^{crit}$ of a few times $10^{4}$, suggesting that the DCBHs might be even rarer than suggested before. Nevertheless, primordial gas with trace amounts of H$_2$ still can form massive objects \citep{latif6}, even if fragmentation is not fully suppressed.

An alternative mechanism to inhibit H$_{2}$ cooling includes the presence of magnetic fields. Magnetic fields are ubiquitous in astrophysical systems and the evidence of their presence at high redshift is growing \citep{bernet,murphy}. However, the origin and strength of these primordial fields is still an open question. A possible explanation of their formation includes astrophysical processes, such as the Biermann battery or the Weibel instability \citep{bierman,schlic}. Even more, it has been proposed that primordial magnetic fields may have appeared during cosmological inflation or through electroweak or quantum chromodynamical phase transitions (see the review by \citet{grasso}). More recent seeding models include the photoionisation process by photons provided by the first luminous sources \citep{Langer2005}, charge segregation ahead of ionization fronts during the Epoch of Reionization, among others \citep{Geraldi2021}. Regardless of their origin, theory predicts the seed fields to be typically weak \citep{subramanian2016}. \citet{Neronov2010} reported a lower limit of $B \gtrsim 3\times10^{-16}$~G and Planck results provide as an upper limit of a nG level at a scale of 1 Mpc \citep{Planckmagnetic,Paoletti2019,Paoletti2022}. Therefore it is of relevant importance to understand how these fields are amplified during structure formation.

The presence of turbulence during the collapse and accretion in the primordial gas has been revealed through several high-resolution simulations of mini- and massive halos, playing an important role in regulating the angular momentum and the fragmentation of the gas \citep{abel,yoshida,Grief,Borm1}. This suggests that in addition to the gravitational compression under the constraint of flux freezing, the amplification of the initial magnetic seed field may be generated by the small-scale dynamo process which transforms the turbulent energy into magnetic energy amplifying the small initial field on very short time-scales \citep{kazantsev,brandenburg}.

\subsection{Implications of the gravitational collapse in a magnetized environment}
Due to the difficulty in achieving the large dynamical range in space and time for an entire system in numerical simulations, previous studies have employed turbulent boxes to study the growth rate of the magnetic field through the small scale dynamo, showing that the magnetic energy grows exponentially \citep[see e.g.][]{Haugen2004,Cho2009,Federrath2011}. On the other hand, in a gravitational collapse environment, where the structures are highly non-linear and not in equilibrium, the signatures of this regime will always be visible. If the magnetic field amplification is due to the compression, the field strength will be directly related to the gas density, however, even for a saturated dynamo, the magnetic field strength should scale with gas density as well, as all relevant energy components such as thermal energy or turbulent kinetic energy will scale with the density during the collapse. Therefore, in this regime the expected results found in turbulent boxes cannot be applied unless high resolution is achieved and therefore the effects of gravity and gravitational collapse have to be taken into account. \citet{Schmidt2013} and \citet{latif7} have shown that in a self-gravitating turbulent gas the magnetic field amplification is well correlated with density; they also found that both the amplification due to small scale dynamo and the amplification due to compression are comparable. The latter is directly related to the numerical resolution of the simulation, where if the turbulence is not sufficiently resolved, the small scale dynamo contribution to the magnetic field amplification might be smaller than the contribution due to compression as the dynamo will not be in an efficient regime. Due to the difficulties in resolving turbulence in numerical simulations, most simulations of gravitational collapse will tend to remain within the kinematic regime, unless the initial field strength is sufficiently high to allow saturation to be reached. The test of possible saturation of the dynamo thus have typically involved the systematic variation of the initial magnetic field strength, essentially looking for a slope in the ratio of the physical magnetic field versus the field expected from pure compression, however even if the dynamo saturation is reached, the amplification by gravitational compression will continue until the system evolves towards a complete saturation \citep{Sur2012}.

\subsection{Resolution studies}
The efficiency of the dynamo is directly related with the numerical resolution of the simulation. Simulations with different resolutions effectively have different numerical viscosities and resistivities and only once the viscosity is sufficiently low, the turbulence in the gas can be resolved and therefore numerically modeled. From small-scale dynamo theory, it is further well-known that the growth rate in the kinematic regime depends on the Reynolds number, defined as $Re = VL/\nu$ with $V$ and $L$ being the characteristic turbulent velocity and length scale, respectively and $\nu$ the viscosity of the gas. This indicates that the growth rate increases with Reynolds number which depends on the numerical resolution \citep{brandenburg,Schober_2012,Bovino2013}. This of course can only continue until saturation occurs, which may depend on the typical Mach numbers in the gas and on the available turbulent energy \citep[see e.g.][]{Federrath2011, Schober2015}.

During gravitational collapse, the Jeans length is the critical scale for driving turbulent motions, so it is of huge importance to well resolve it, specially when the dynamo operates. In the context of Pop III star formation in minihalos \citet{Sur2010} performed the first systematic study about the implications of varying the resolution in the behaviour of turbulent gas during collapse and how it interacts with the magnetic field. It was found that the vorticity in the gas increases in simulations of high numerical resolution, as well as the magnetic field strength. They found an additional amplification beyond compression in simulations with a resolution of 64 cells per Jeans length. \citet{FederrathSur2011} performed a Fourier analysis of the magnetic energy in these simulations and found that the spectra of these magnetic fields show the typical signatures that are present in the small-scale dynamo for sufficiently high resolution to resolve the turbulence, i.e. more than 30 cells per Jeans length. Moreover, \citet{Turk2012} and \citet{latif7} performed a similar study focusing on the evolution of the magnetic field during the gravitational collapse in cosmological simulations, finding that the amplification beyond compression is reduced due to the more diffusive numerical scheme employed suggesting that a higher Jeans resolution of $\geq 64$ cells is required. In addition, \citet{Turk2012} reported for minihalos that when using a resolution of 64 cells per Jeans length the magnetic field strength increases by 7 orders of magnitude over the effect of compression while the density shows an increment of $16$ orders of magnitude. On the other hand \citet{latif7} reported that in atomic cooling halos when employing a resolution of $128$ cells per Jeans length the amplification due to the small scale dynamo is 2 orders of magnitude above the compression, decreasing when using a smaller resolution. This illustrates the  challenge of capturing and identifying such effects in numerical simulations.

\subsection{Magnetic field effect in the fragmentation of primordial clouds}
It has been shown that the small-scale dynamo in atomic cooling halos can efficiently amplify the magnetic field into a saturated state in the presence of strong accretion shocks, thereby helping in suppressing fragmentation via additional magnetic pressure \citep{Schleicher2010,latif7,latif8}. Recently, \citet{grete2019} included for the first time a subgrid scale (SGS) model for unresolved MHD turbulence to explore its impact on the formation of DCBHs, finding that the fragmentation during the collapse is intermittent with accretion rates sufficiently large to support the direct collapse scenario. 

Until recently, these studies explored how the magnetic fields were generated, amplified and how they evolved across different scales, but due to numerical constraints, it was difficult to study how the magnetic fields affected the initial mass function of these objects. \citet{sharda1,sharda2} answered this question in the context of the so-called minihalos, showing that firstly an initial weak magnetic field can grow via small-scale and large-scale mean-field dynamos and secondly, that the fragmentation differs significantly from simulations without magnetic fields, therefore concluding that magnetic fields have a significant impact on the primordial IMF. This marks the starting point of further investigations in this area, as the impact of magnetic fields in the larger atomic cooling halos is not yet understood.

\citet{Hirano2021} studied the effect of the magnetic field on star formation in atomic-cooling halos focusing on the early accretion phase of the atomic gas cloud. They showed that the magnetized atomic gas clouds fragment to a number of dense cores which merge intermittently into the most massive core due to the transfer of angular momentum by the magnetic field. Moreover, ignoring the effect of turbulence, they showed that the magnetic field is efficiently amplified by the motion of the dense cores. Similarly, \citet{Latif2022} conducted cosmological simulations and evolved them for longer time finding that in the MHD case, the initial clump masses are higher but the fragmentation is reduced because the disks are more stable. In contrast, in the non-MHD case, the initial clump mass is lower, but the higher merger rate observed due to the fragmentation yields similar masses of the central clump compared to the MHD case. They did not include H$^-$ cooling in their chemical network, which, as shown in \cite{latif2016}, causes small-scale fragmentation. Therefore, they suggest that future simulations should include this process. \citet{Hirano2022} explored the effect of metals in magnetized atomic cooling halos (though without resolving smaller-scale turbulence and therefore the small-scale dynamo process). They found that although increasing the metallicity reduces mass accretion, many protostars form in the collapsing central region, which is gravitationally and thermally unstable. This leads to an increase in the magnetic field and promotes subsequent gas accretion and coalescence of the low-mass protostars regardless of the initial magnetic field strength. These findings imply that magnetic fields can reduce some of the requirements for the direct collapse scenario and further investigation with more realistic cosmological scenarios by including also the effect of molecular cooling is needed.

In this paper, we explore the evolution of magnetic fields in halos with $\gtrsim10^7$~M$_\odot$ collapsing at redshifts $z\gtrsim12$ under different conditions, particularly for different strengths of the Lyman-Werner background corresponding to the regimes of atomic and molecular hydrogen cooling, including different initial values of the magnetic field strength. We present our methodology in section~\ref{methods}, the results in section~\ref{results} and our summary and discussions in section~\ref{summary}.

\section{Computational Methods}\label{methods}

We conduct cosmological magneto-hydrodynamical (MHD) zoom-in simulations by using the 3D MPI-parallel, Eulerian, block structured, adaptive mesh refinement (AMR) code ENZO \citep{Bryan2014,enzo2}. This code is open-source and was designed for self-graviting compressible fluid dynamics, including the effects of radiative transfer, magnetic fields and multiple subgrid and microphysical processes. The main physical equations solved by the code are the Eulerian equation of cosmological (comoving) ideal MHD including gravity. Additionally, the Poisson equation for the gravitational potential is solved. In order to solve the MHD equations we employed the Dedner method which is based on the Godunov MUSCL scheme and cleans the divergence constraint $\nabla\cdot B=0$ with a wave-like hyperbolic cleaner \citep{Dedner2002}. For the reconstruction of  the variables we used the piecewise linear method (PLM) \citep{vanleer1979} and to solve the Riemann problem we used the Harten-Lax-van Leer (HLL) Riemann solver \citep{Toro1997}. It is important to note that the Riemann solver used here is somewhat more diffusive compared to the HLL3R scheme used in similar works with the Flash code \citep[see e.g.][]{Sur2010,FederrathSur2011,Sur2012}, therefore the required Jeans resolution to resolve the turbulence can be larger \citep{Turk2012,latif7}.

\subsection{Initial conditions}
To perform cosmological zoom-in simulations, we followed the approach of \cite{grete2019}. We generated our cosmological nested grid initial conditions with MUSIC \citep{Hahn2011} which uses Lagrangian perturbation theory to obtain initial velocity and displacements fields based on numerical solutions of Poisson's equation. Additionally the density perturbations are generated through Gaussian random fields that follow a prescribed power spectrum. For the cosmological parameters we use the following data provided by \cite{planck2016}: $\Omega_{m}=0.3089$, $\Omega_{b}= 0.048598$, $\Omega_{\Lambda}= 0.6911$, $\sigma_{8}= 0.8159$, $h=0.677$ and $n=0.9667$.

We start with a unigrid dark matter (DM) only simulation within a computational box with a side length of $1$~Mpc~$h^{-1}$  at $z=100$ using a top grid resolution of $1024^3$ cells. We evolved the simulations until $z=12$ and identify the most massive halo using the Rockstar halo finder \citep{rockstar} to then trace back the DM particles position in the Lagrangian volume of the halo within a region of two times its virial radius at the initial redshift. After this procedure we were able to generate new nested initial conditions, therefore we reran the simulations including baryons and additional physics using a top grid resolution of $256^3$ cells. Two additional nested refined levels each with the same resolution were employed yielding an effective spatial resolution of $1024^3$ cells, identically to our DM only simulations. We repeated this process three times varying our initial conditions by using different random seeds in MUSIC, which allowed us to analyse three different halos. From this, we were able to reach a maximum DM mass resolution of $99$~M$_{\sun}$ with $\sim2\times10^7$ DM particles on the halos. We further added 22 to 26 levels of refinement that allow us to reach a density peak of about $3\times10^{-13}$~g/cm$^{3}$. This is the density at which we stop all our simulations, therefore our spatial resolution ranges between of $19.3$~au to $2.2$~au depending on the halo. In this work a grid is refined if one of the following criteria is triggered: DM overdensity of a factor of 4, baryon overdensity of a factor of 4 with a refinement level exponent of $-0.3$ which makes the refinement super-Lagrangian and the Jeans length resolution where we employed 4 different values, 32, 64, 128 and 256 cells. Such higher values are used to resolve the turbulent eddies for which small-scale dynamo gets excited and may exponentially amplify magnetic fields as found in previous studies \citep{latif7,grete2019}. We smoothed the DM particles at the refinement level 12 to avoid numerical artefacts; this corresponds to 1.4~pc in comoving units. Furthermore, we have started our simulations with three uniform magnetic field seeds, each with a proper strength of $10^{-14}$~G, $10^{-10}$~G and $10^{-8}$~G. According to \cite{Planckmagnetic} the upper limits for the primordial magnetic field in the CMB is a few comoving nG at a scale of 1 Mpc and the values used in this work for the initial magnetic field are below that limit ($10^{-9}$~nG, $10^{-5}$~nG and $10^{-3}$~nG in comoving units).

\subsection{Chemical model}
We solved the chemical and thermal evolution of the gas using the open-source chemistry package KROME \citep{krome}. 
We used the primordial chemical model presented in \cite{latif2016} which solves the rate equations of 9 different chemical species ($\mathrm e^-$, $\mathrm H^-$, H, $\mathrm H^+$, He, He$^+$, He$^{++}$, $\mathrm H_2$, $\mathrm H_2^+$) self-consistenly with the MHD simulations. Among the relevant processes that this model includes are chemical heating, chemical cooling, atomic cooling, H$_2$ cooling, H$_2$ self-shielding, H$_2$ photodissociation heating and H$^-$ photodetachment heating. It also includes H$^-$ cooling at higher densities. Additionally, based on \citet{glover1,glover2} we include extra chemical reactions for completeness such as collisional ionization of atomic hydrogen due to H–H collisions and collisions with neutral helium. Chemical species related to deuterium are not considered in the model as they are relevant at lower temperatures and get easily dissociated with lower LW fluxes. We assume a uniform UV background and varied its strength from $J_{21}=10^2$ to $J_{21}=10^5$ using black-body shaped spectra with temperature of $T_{rad}=2\times10^4$~K to explore the amplification of magnetic fields in different environments including the effect of the molecular hydrogen cooling. 

\section{Results}\label{results}

\begin{table}
\caption{Properties of the simulated halos using an initial magnetic field seed strength of $B_0=10^{-14}$ G (proper). It includes the name of the halo, the Lyman-Werner strength, the resolution per Jeans length, the magnetic field at the centre of the halo taken from the mass-weighted binned average profile, the mass of the halo and the redshift when the halo reaches a density peak of $\sim 3\times10^{-13}$ g/cm$^{3}$.}             
\resizebox{\columnwidth}{!}{
\label{tab:table1}      
\centering                          
\begin{tabular}{c c c c c c}        
\hline\hline                 
Halo & $\mathrm{J}_{21}$ & Jeans res. & $\mathrm{B}_c$ [G] & $\mathrm{M}_{halo}$ [$\mathrm{M}_\odot$] & $z$\\    
\hline                        
1  & $10^2$ & 32 cells & $2.63\times10^{-7}$ & $1.80\times10^{7}$&15.382\\
1  & $10^2$ & 64 cells & $1.11\times10^{-5}$ & $1.85\times10^{7}$&15.302\\
1  & $10^2$ & 128 cells & $2.76\times10^{-6}$ & $2.17\times10^{7}$ &15.022\\
1  & $10^3$ & 32 cells & $4.14\times10^{-7}$ & $4.47\times10^{7}$&14.082\\
1  & $10^3$ & 64 cells & $7.25\times10^{-7}$ & $4.30\times10^{7}$& 14.152\\ 
1  & $10^3$ & 128 cells & $5.15\times10^{-7}$ & $4.33\times10^{7}$&14.139\\
1  & $10^3$ & 256 cells & $1.73\times10^{-6}$ & $4.52\times10^{7}$ &14.165\\
1  & $10^4$ & 32 cells & $4.46\times10^{-7}$ & $4.53\times10^{7}$&14.057\\
1  & $10^4$ & 64 cells & $5.81\times10^{-7}$ & $4.50\times10^{7}$&14.064\\
1  & $10^4$ & 128 cells & $1.19\times10^{-5}$ & $4.45\times10^{7}$ &14.082\\
1  & $10^5$ & 32 cells & $5.05\times10^{-7}$ & $4.53\times10^{7}$&14.056\\
1  & $10^5$ & 64 cells & $9.61\times10^{-8}$ & $4.50\times10^{7}$&14.062\\
1  & $10^5$ & 128 cells & $3.08\times10^{-6}$ &  $4.46\times10^{7}$&14.082\\
1  & $10^5$ & 256 cells & $5.60\times10^{-5}$ & $4.58\times10^{7}$ &14.139\\
\hline
2  & $10^2$ & 128 cells & $9.12\times10^{-7}$ &$3.17\times10^{7}$ &13.669\\
2  & $10^3$ & 32 cells & $1.34\times10^{-6}$ &$5.93\times10^{7}$ &12.589\\
2  & $10^3$ & 64 cells & $1.28\times10^{-7}$ & $5.96\times10^{7}$&12.579\\
2  & $10^3$ & 128 cells & $1.03\times10^{-4}$ & $5.73\times10^{7}$ &12.616\\
2  & $10^4$ & 128 cells & $1.16\times10^{-4}$ &$5.96\times10^{7}$ &12.569\\
2  & $10^5$ & 32 cells & $1.98\times10^{-7}$ & $6.09\times10^{7}$&12.516\\
2  & $10^5$ & 64 cells & $4.98\times10^{-6}$ & $6.13\times10^{7}$&12.522\\
2  & $10^5$ & 128 cells & $2.80\times10^{-5}$ & $5.96\times10^{7}$ &12.564\\
\hline
3  & $10^2$ & 32 cells & $7.84\times10^{-6}$ & $2.42\times10^{7}$ &14.821\\
3  & $10^2$ & 64 cells & $2.45\times10^{-7}$ & $2.39\times10^{7}$&14.869\\
3  & $10^2$ & 128 cells & $1.84\times10^{-5}$ & $2.41\times10^{7}$&14.879\\
3  & $10^3$ & 32 cells & $5.45\times10^{-7}$ & $4.09\times10^{7}$&13.784\\
3  & $10^3$ & 64 cells & $1.27\times10^{-6}$ & $4.20\times10^{7}$&13.712\\
3  & $10^3$ & 128 cells & $1.00\times10^{-5}$ & $4.24\times10^{7}$ &13.693\\
3  & $10^4$ & 32 cells & $5.82\times10^{-7}$ & $4.32\times10^{7}$&13.627\\
3  & $10^4$ & 64 cells & $8.09\times10^{-6}$ & $4.35\times10^{7}$ &13.600\\
3  & $10^4$ & 128 cells & $1.61\times10^{-5}$ & $4.42\times10^{7}$ &13.567\\
3  & $10^5$ & 32 cells & $3.79\times10^{-7}$ & $4.31\times10^{7}$ &13.628\\
3  & $10^5$ & 64 cells & $4.96\times10^{-6}$ & $4.35\times10^{7}$&13.604\\
3  & $10^5$ & 128 cells & $1.23\times10^{-5}$& $4.41\times10^{7}$&13.572\\
\hline                                   
\end{tabular}
}
\end{table}

In the following, we present our results concerning magnetic field amplification and saturation in more complex primordial cooling scenarios under the influence of Lyman-Werner radiation. In subsection~\ref{high} we present results regarding the amplification and saturation of the magnetic field by using different initial magnetic field seeds and different Jeans resolutions, while in subsection~\ref{UV} we focus particularly on the effect resulting from the strength of the Lyman-Werner background. We simulated three different halos but as they presented similar results, here we based the analysis on halo 1 which is the one that includes our highest Jeans resolution simulations. The radial profiles in this section show mass-weighted average quantities (for comparison with other statistical methods see Appendix~\ref{statistic}) and all of them are presented in proper units. In addition, the results and physical/magnetic properties for the other halos are provided in Appendix~\ref{Extrahalos}.


\subsection{Amplification and saturation of the magnetic field}\label{high}
   \begin{figure*}
   \centering
   \includegraphics[width=0.95\textwidth]{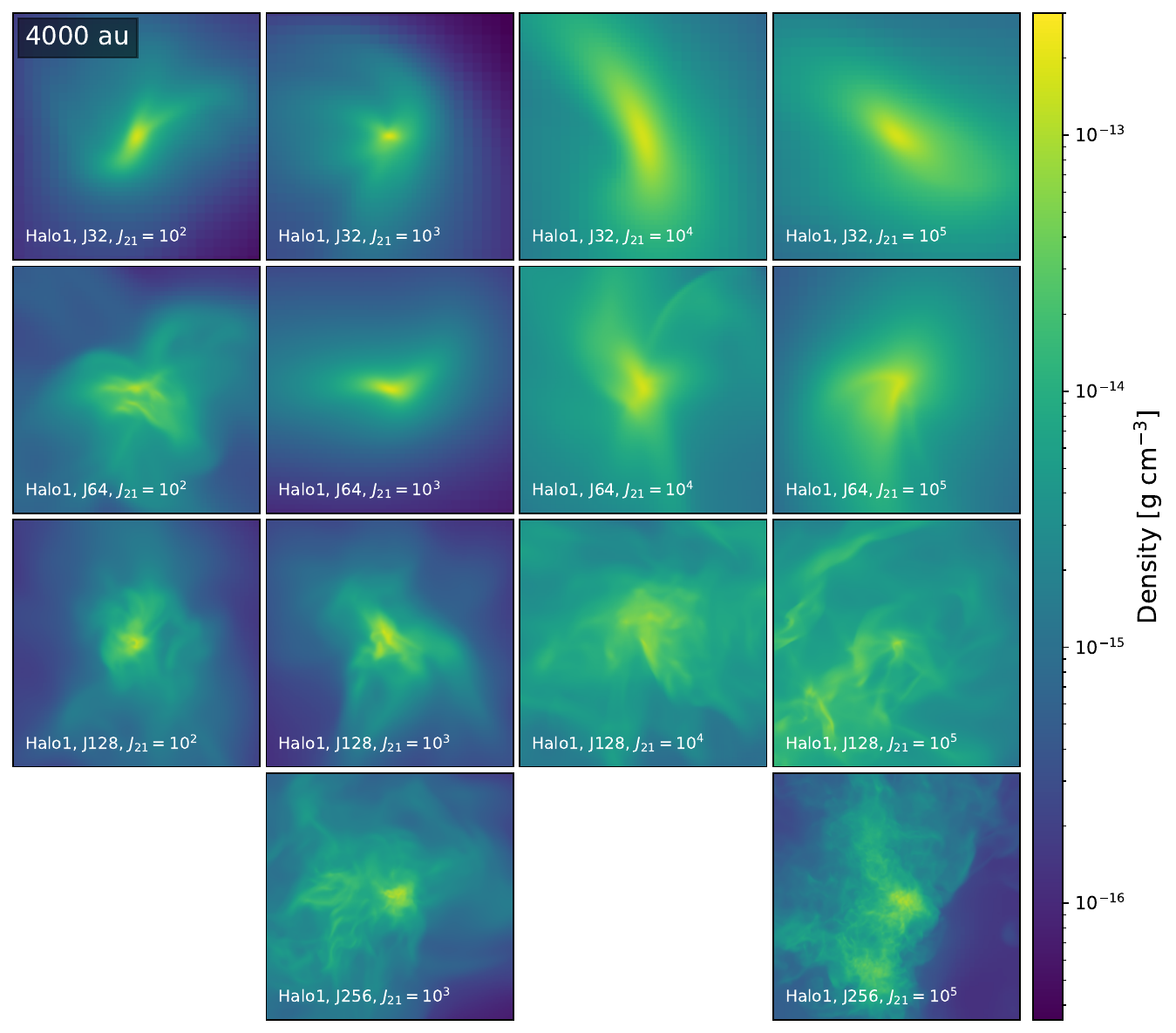}
    \caption{Density-weighted projections of the density along the z-axis with a physical width of 4000~au for halo 1 when reaching a peak density of $3\times10^{-13}$ $\mathrm{[g\,cm^{-3}]}$ using an initial magnetic field of $B_0=10^{-14}\,\mathrm{[G]}$ (proper). From top to bottom we vary the Jeans resolution using 32, 64, 128 and 256 cells per Jeans length and from left to right we vary the strength of the radiation background to $J_{21}=10^2$, $10^3$, $10^4$ and $10^5$. The halo shows more complex structures with increasing the Jeans resolution, as the turbulence seems to be better resolved. This occurs regardless of the value of $J_{21}$.}
    \label{fig:densityproj}
    \end{figure*}

   \begin{figure*}
   \centering
   \includegraphics[width=0.95\textwidth]{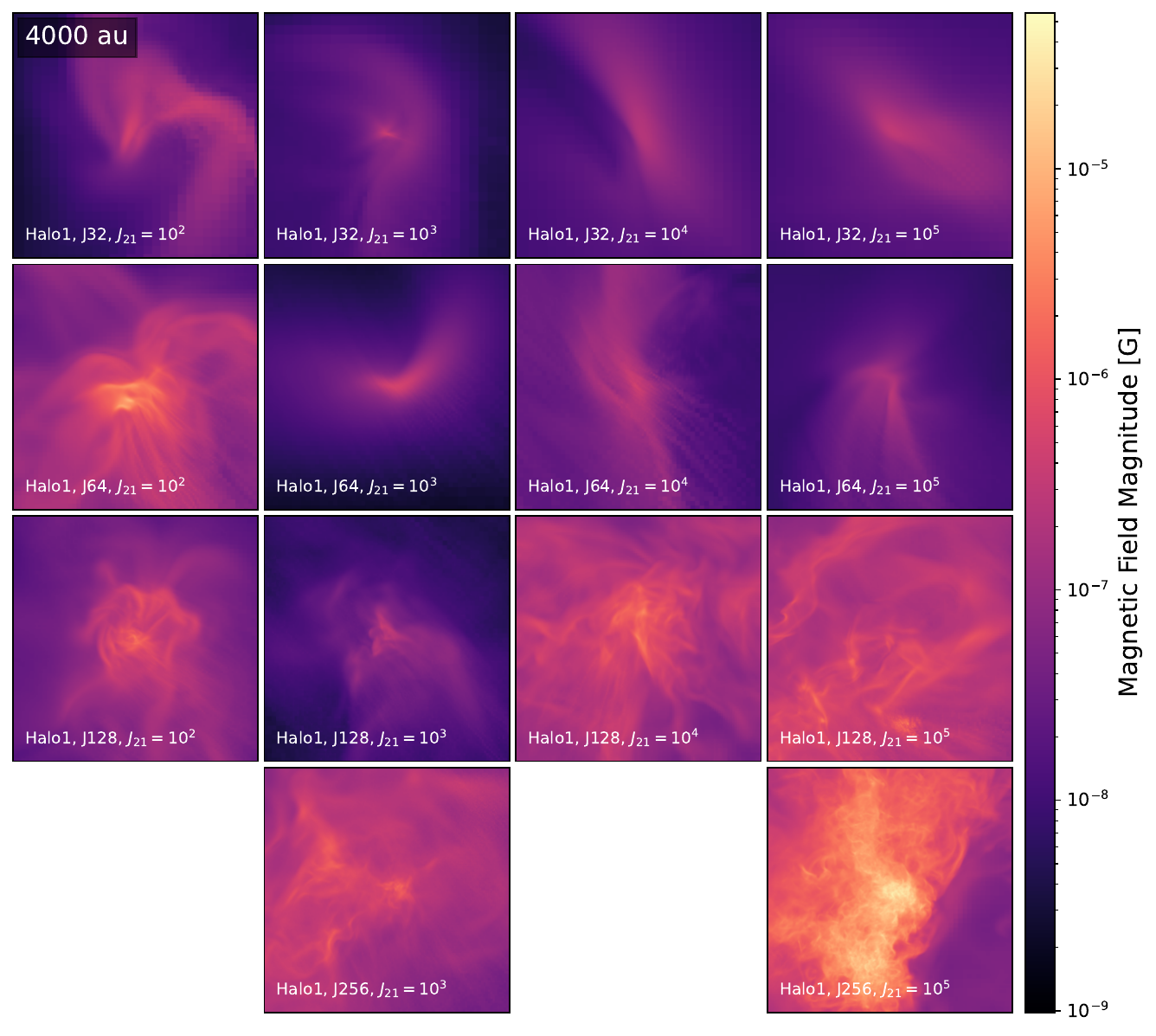}
    \caption{Density-weighted projection of magnetic field magnitude along the z-axis with a physical width of 4000~au for halo 1 when reaching a peak density of $3\times10^{-13}$ $\mathrm{[g\,cm^{-3}]}$ using an initial magnetic field of $B_0=10^{-14}\,\mathrm{[G]}$ (proper). From top to bottom we vary the Jeans resolution using 32, 64, 128 and 256 cells per Jeans length and from left to right we vary the strength of the radiation background to $J_{21}=10^2$, $10^3$, $10^4$ and $10^5$. As the Jeans resolution increases, the structure of the magnetic field starts to become more complex and begins to expand outwards, filling more volume in the halo.}
   \label{fig:bproj}
    \end{figure*}

   \begin{figure*}
   \centering
   \includegraphics[width=0.95\textwidth]{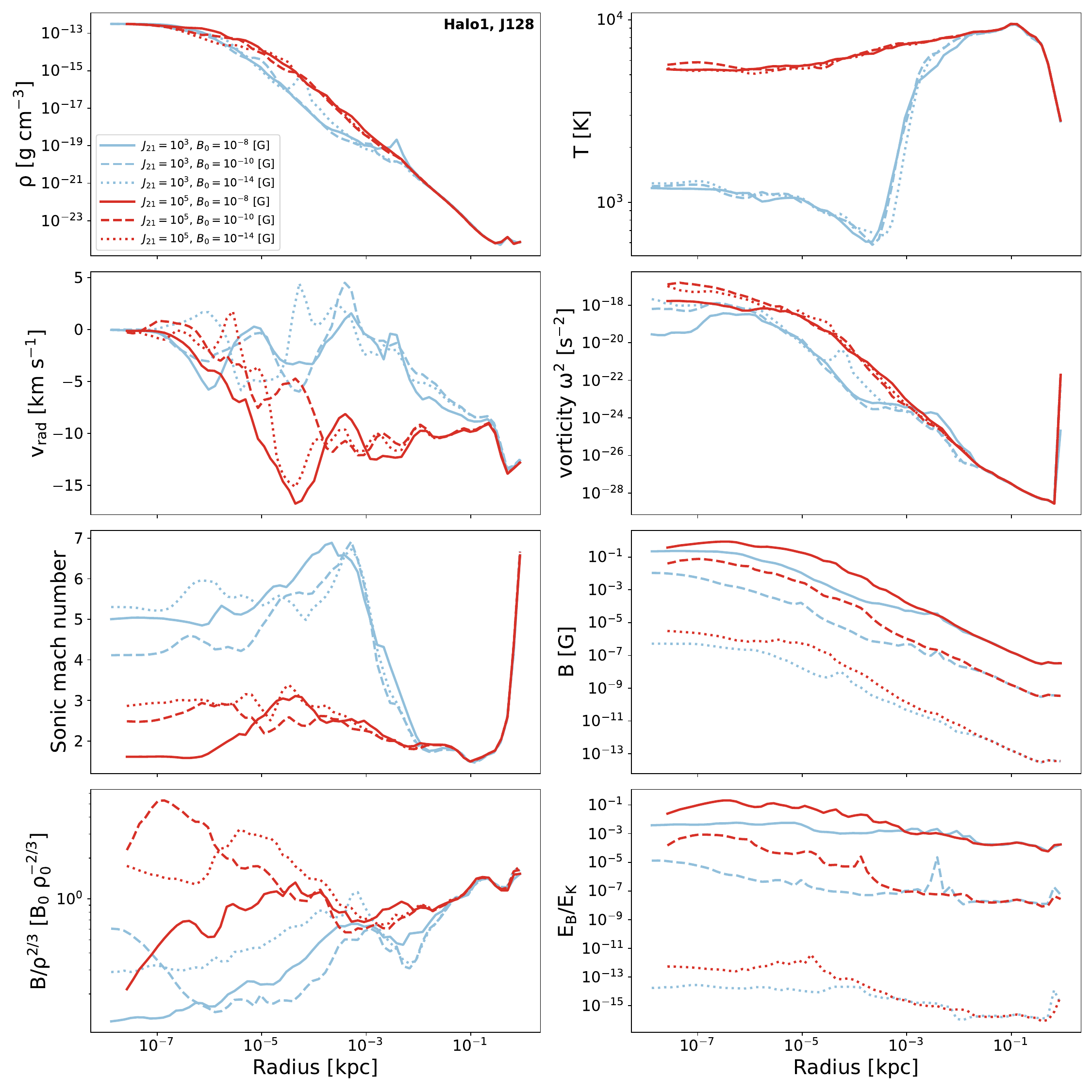}
   \caption{Mass-weighted spherically binned radial profiles of density, temperature, radial velocity, vorticity squared, sonic Mach number, magnetic field strength, magnetic field amplification $B/\rho^{2/3}$ and the magnetic to kinetic energy density ratio $E_B/E_K$ for halo 1 when reaching a peak density of $3\times10^{-13}$ $g/cm^{3}$ using a fixed Jeans resolution of 128 cells per Jeans length. The quantity $B/\rho^{2/3}$ is normalized by  $B_0/\rho_0^{2/3}$ with $B_0$ the initial magnetic field strength and $\rho_0$ the cosmic mean baryon density at $z=100$. Light blue lines represent the simulations with $J_{21}=10^3$ where the cooling is driven via molecular hydrogen and red lines are for simulations with $J_{21}=10^5$ where the cooling is driven by atomic hydrogen. Solid lines are for runs with $B_0=10^{-8}$~[G] (proper), dashed lines for runs with $B_0=10^{-10}$~[G] (proper) and dotted lines for runs with $B_0=10^{-14}$~[G] (proper). Infall velocity and vorticity tends to be larger for the simulations using $J_{21}=10^5$ while the sonic mach number tends to be smaller compared to the simulations with $J_{21}=10^3$ following the thermal evolution of the halo. Magnetic field strength, $B/\rho^{2/3}$ and $E_B/E_K$ also tends to be larger for the simulations in the atomic cooling regime which is consistent with the behaviour of the physical properties of the halo.}
    \label{fig:structhalo3}
    \end{figure*}

   \begin{figure*}
   \centering
   \includegraphics[width=\textwidth]{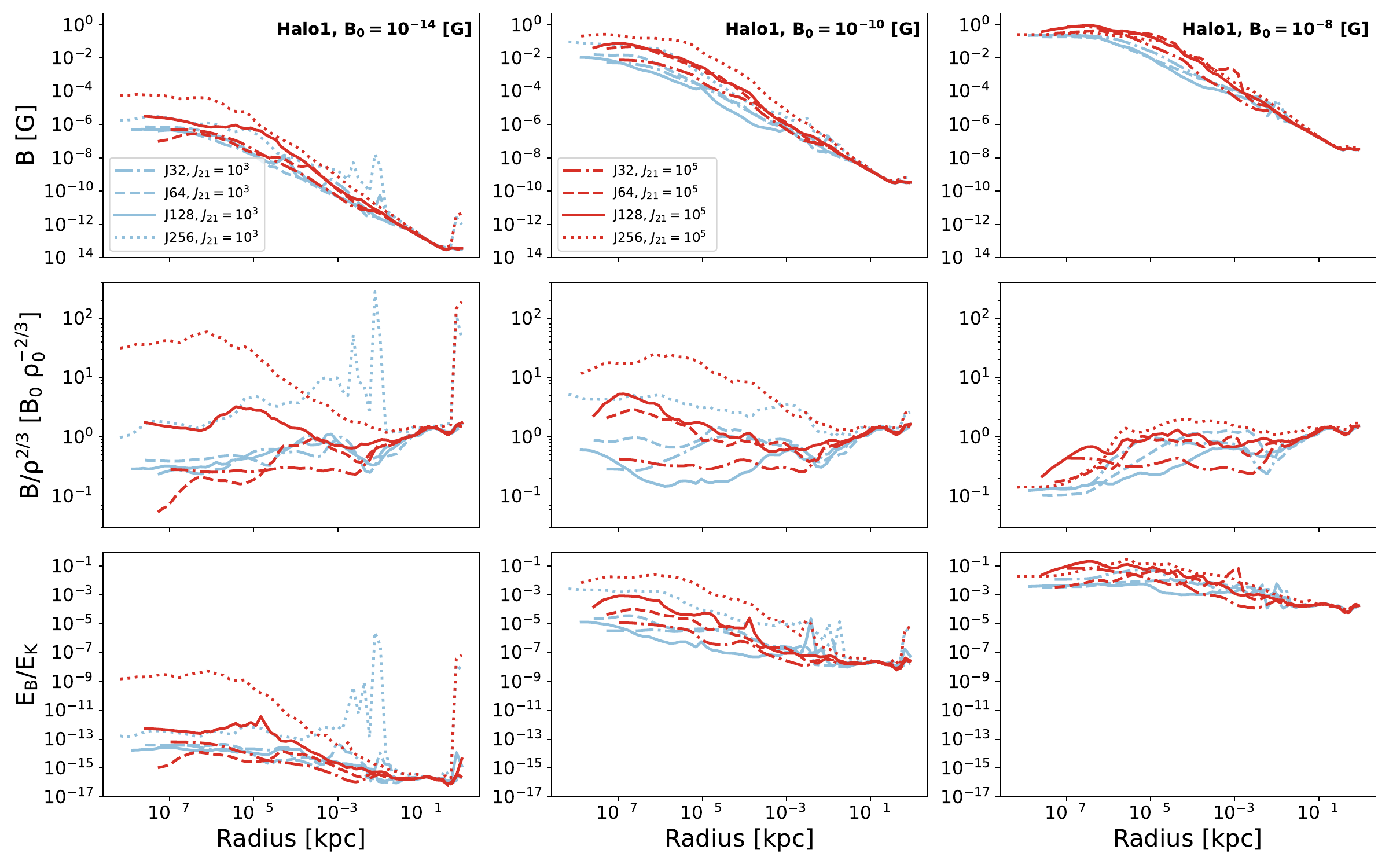}
   \caption{Mass-weighted spherically binned radial profiles of magnetic field strength, magnetic field amplification $B/\rho^{2/3}$ and the magnetic to kinetic energy density ratio $E_B/E_K$ for halo 1 when reaching a peak density of $3\times10^{-13}$ $g/cm^{3}$. The first, second and third column of this multiple plot shows the magnetic properties of the runs with a proper initial magnetic field strength of $B_0=10^{-14}$~[G], $B_0=10^{-10}$~[G] and $B_0=10^{-8}$~[G] respectively. Light blue lines represent the simulations with $J_{21}=10^3$ where the cooling is driven via molecular hydrogen and red lines are for simulations with $J_{21}=10^5$ where the cooling is driven by atomic hydrogen. The different line styles represent different Jeans resolutions; dash-dotted line for 32 cells, dashed line for 64 cells, solid line for 128 cells and dotted line for 256 cells per Jeans length. Independently of the value of $J_{21}$, the magnetic field strength, as well as the other magnetic properties, increases with increasing the Jeans resolution for the runs with $B_0=10^{-14}$~[G]. When increasing the initial magnetic field the simulations with $J_{21}=10^3$ start to present a dependency loss on the Jeans resolution at lower $B_0$ in comparison with the atomic cooling runs which suggest that potentially the saturation occurs first in these simulations.}
    \label{fig:bhalo3}
    \end{figure*}

   \begin{figure*}
   \centering
   \includegraphics[width=\textwidth]{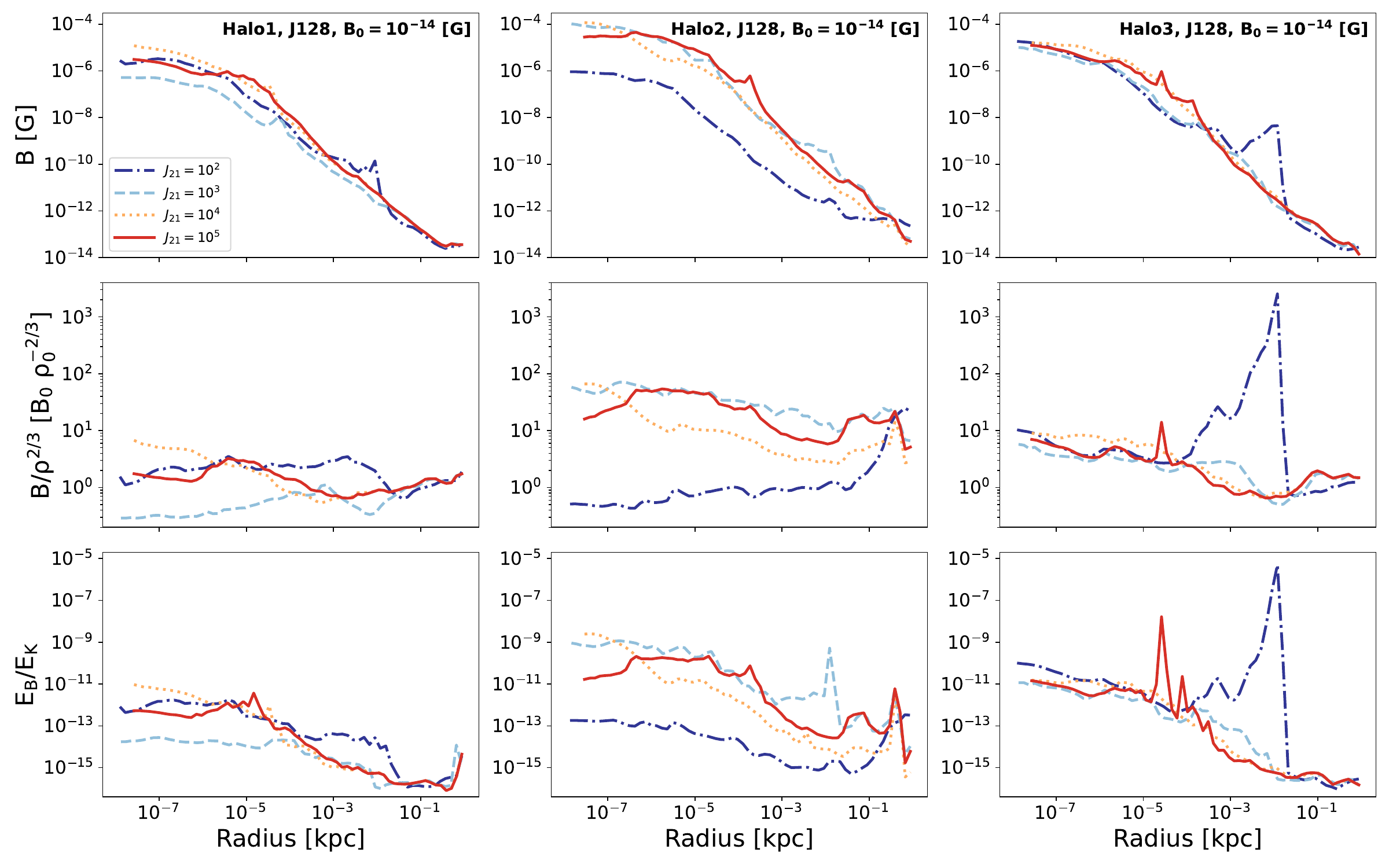}
    \caption{Mass-weighted spherically binned radial profiles of magnetic field strength, magnetic field amplification $B/\rho^{2/3}$ and the magnetic to kinetic energy density ratio $E_B/E_K$ for the three different halos when reaching a peak density of $3\times10^{-13}$ $g/cm^{3}$ using a fixed Jeans resolution of 128 cells and $B_0=10^{-14}$~[G] (proper). The first, second and third column of this multiple plot shows the magnetic properties of halo 1, halo 2 and halo 3 respectively. The dash-dotted blue line is for $J_{21}=10^2$, the dashed light blue line is for $J_{21}=10^3$, the dotted orange line is for $J_{21}=10^4$ and the solid red line is for $J_{21}=10^5$. The effect of the UV background on the amplification of the magnetic field can vary from halo to halo, however, independently of this value, the magnetic field is amplified by 
    $\sim 7- 10$ orders of magnitude.}
    \label{fig:bhalo3UV}
    \end{figure*}
\begin{table}
\caption{Same as Table~\ref{tab:table1} but for a higher initial magnetic field seed strength of $B_0=10^{-10}$ G (proper).}             
\resizebox{\columnwidth}{!}{
\label{tab:table3}      
\centering                          
\begin{tabular}{c c c c c c}        
\hline\hline                 
Halo & $\mathrm{J}_{21}$ & Jeans res. & $\mathrm{B}_c$ [G] & $\mathrm{M}_{halo}$ [$\mathrm{M}_\odot$] & $z$\\    
\hline                        
1  & $10^3$ & 32 cells & $5.12\times10^{-3}$ & $4.47\times10^{7}$ &14.085\\
1  & $10^3$ & 64 cells & $1.56\times10^{-2}$ & $4.30\times10^{7}$ & 14.160\\
1  & $10^3$ & 128 cells & $1.07\times10^{-2}$ & $4.33\times10^{7}$ & 14.138\\
1  & $10^3$ & 256 cells & $9.29\times10^{-2}$ & $4.51\times10^{7}$ &14.171\\
1  & $10^5$ & 32 cells & $7.55\times10^{-3}$ & $4.53\times10^{7}$ &14.059\\
1  & $10^5$ & 64 cells & $3.71\times10^{-2}$ & $4.50\times10^{7}$&14.066\\
1  & $10^5$ & 128 cells & $4.05\times10^{-2}$ & $4.45\times10^{7}$ &14.085\\
1  & $10^5$ & 256 cells & $2.06\times10^{-1}$ &  $4.57\times10^{7}$&14.141\\
\hline
3  & $10^3$ & 32 cells & $3.64\times10^{-2}$ & $4.07\times10^{7}$ &13.797\\
3  & $10^3$ & 64 cells & $7.74\times10^{-3}$ & $4.18\times10^{7}$ &13.723\\
3  & $10^3$ & 128 cells & $2.41\times10^{-2}$ & $4.22\times10^{7}$ &13.716\\
3  & $10^5$ & 32 cells & $6.02\times10^{-3}$  & $4.32\times10^{7}$ &13.625\\
3  & $10^5$ & 64 cells & $1.61\times10^{-2}$ & $4.35\times10^{7}$ &13.602\\
3  & $10^5$ & 128 cells &$1.34\times10^{-1}$ & $4.41\times10^{7}$&13.573\\
\hline                                   
\end{tabular}
}
\end{table}

\begin{table}
\caption{Same as Table~\ref{tab:table1} but for a higher initial magnetic field seed strength of $B_0=10^{-8}$ G (proper).}             
\resizebox{\columnwidth}{!}{
\label{tab:table4}      
\centering                          
\begin{tabular}{c c c c c c}        
\hline\hline                 
Halo & $\mathrm{J}_{21}$ & Jeans res. & $\mathrm{B}_c$ [G] & $\mathrm{M}_{halo}$ [$\mathrm{M}_\odot$] & $z$\\    
\hline                        
1  & $10^3$ & 32 cells & $2.44\times10^{-1}$ & $4.47\times10^{7}$ &14.083\\
1  & $10^3$ & 64 cells & $1.84\times10^{-1}$ & $4.39\times10^{7}$ & 14.114\\
1  & $10^3$ & 128 cells & $2.23\times10^{-1}$ & $4.38\times10^{7}$ &14.119\\
1  & $10^5$ & 32 cells & $7.76\times10^{-1}$ & $4.53\times10^{7}$ &14.059\\
1  & $10^5$ & 64 cells & $3.07\times10^{-1}$ & $4.53\times10^{7}$ &14.052\\
1  & $10^5$ & 128 cells & $3.80\times10^{-1}$ & $4.46\times10^{7}$&14.076\\
1  & $10^5$ & 256 cells & $2.51\times10^{-1}$ & $4.59\times10^{7}$&14.139\\
\hline
3  & $10^3$ & 32 cells & $3.63\times10^{-1}$  & $4.08\times10^{7}$ & 13.802\\
3  & $10^3$ & 64 cells & $2.40\times10^{-1}$  & $4.20\times10^{7}$&13.713\\
3  & $10^3$ & 128 cells & $1.81\times10^{-1}$ & $4.22\times10^{7}$ &13.716\\
3  & $10^5$ & 32 cells & $1.22$ & $4.31\times10^{7}$ &13.628\\
3  & $10^5$ & 64 cells & $2.55\times10^{-1}$ &$4.35\times10^{7}$ &13.601\\
3  & $10^5$ & 128 cells &$5.31\times10^{-1}$ & $4.38\times10^{7}$&13.595\\
\hline                                   
\end{tabular}
}
\end{table}

We start presenting the simulations for halo 1 pursued for different Lyman-Werner backgrounds and varying the resolution per Jeans length. The properties of the three simulated halos for $B_0=10^{-14}$~G (proper) are listed in Table~\ref{tab:table1}. In addition, as we aim to at least obtain an approximate idea for which magnetic field strength the dynamo saturates and no further dynamo amplification occurs (though compression via collapse may continue), we have pursued simulations where we adopted a higher initial magnetic field strength. We study the saturation level of magnetic fields by increasing the initial magnetic field strength due to the difficulty in achieving high Reynolds numbers in numerical simulations. Otherwise we will underestimate their growth rate \citep{Schober2012PhysRev}. A list of the properties of these simulations is given in Table~\ref{tab:table3} for $B_0=10^{-10}$~G and in Table~\ref{tab:table4} for $B_0=10^{-8}$~G. These simulations were evolved until they reach a peak density of about $3\times10^{-13}$~g/cm$^3$, which corresponds to a maximum refinement level of $22-26$ (depending on the initial conditions).

Fig.~\ref{fig:densityproj} shows the density-weighted projection of the density in the central 4000~au of halo 1, using an initial magnetic field seed of $B_0=10^{-14}$~G (proper). From visual inspection, it is clearly recognizable that the density structure in the central region changes considerably with the Jeans resolution. For 32 cells per Jeans length we cannot see turbulent structures in the central part of the halo and the structures appear rather compact, sometimes elongated and sometimes more round. With a Jeans resolution of 64 cells, the turbulence seems to be better resolved and more complex structures start to become visible. This effect becomes more pronounced in case of 128 cells per Jeans length, where turbulent filamentary structures fill a significant part of the central volume. For 256 cells per Jeans length the turbulence is even better resolved, where the complex structures appear in more detail compared to the previous cases. These trends in principle occur for all of the values of the Lyman-Werner background that we have considered. Even though we start to resolve some turbulence using a resolution of 64 cells per Jeans length as suggested by previous works \citep[see e.g.][]{Turk2012,latif7,grete2019}, we find that actually more complex structures start to appear in this particular halo with a Jeans resolution of $\geq128$ cells reflecting the challenge of resolving turbulence in numerical simulations.

In Fig.~\ref{fig:bproj} we show the projection of the magnetic field in the central region of the halo. In the runs with 64 cells per Jeans length, the magnetic field is centrally concentrated or in elongated filaments, while for the higher resolutions per Jeans length, we clearly see how the magnetic field structure becomes more volume filling, the magnetic field strength increases, and its structures become more complex due to the interaction with the underlying velocity field. This effect seems to be rather independent of the adopted value of $J_{21}$.

For a more quantitative analysis of the structure of the halos and its physical properties, we present in Fig.~\ref{fig:structhalo3} the radial profiles of the density, temperature, radial velocity, vorticity squared and sonic Mach number for halo 1 for a subset of simulations, focusing on $J_{21}=10^3$ and $J_{21}=10^5$ and on the different initial magnetic field seeds for comparison. In this plot we fixed the Jeans resolution to 128 cells. The density profile is almost independent of the initial magnetic field seed and the adopted Lyman-Werner background and approximately follows the profile of an isothermal sphere in the outer parts of the halo, while it becomes approximately flat at the centre. At a scale roughly of the order $10^{-4}$~kpc, a small bump is present in the $B_0=10^{-14}$~G run for $J_{21}=10^3$ due to inhomogeneities in the density field. This also happens in our strongest initial magnetic field run for $J_{21}=10^3$ in which a bump is present at a scale between $10^{-3}-10^{-2}$~kpc.

The radial profile of temperature is the same for all runs at least on large scales where it is driven by the virialization shock and chemistry at low densities where molecular hydrogen is fully dissociated in all simulations. From a scale of roughly $10^{-2}$~kpc, the behaviour however starts to diverge, and simulations with $J_{21}=10^5$ remain at a high temperature of $\sim8000$~K due to H$_2$ photodissociation until it reaches a scale of about $10^{-3}$~kpc where the gas cools down to $5000-6000$~K keeping these temperatures even on smaller scales due to the presence of H$^-$ cooling, as it was found in \citet{latif2016}, while for $J_{21}=10^3$ molecular hydrogen efficiently forms and the gas cools, leading to a temperature minimum of $\sim500$~K on a scale of $\sim10^{-3}$~kpc with subsequent moderate increase towards smaller scales, as the de-excitation due to collisions becomes more important at higher densities compared to radiative losses. The difference in the temperature also leads to a moderate difference in the density on the scales below $10^{-2}$~kpc, where the density is higher for simulations with $J_{21}=10^5$ compared to $J_{21}=10^3$. Overall, one should also stress that the changes in temperature are moderate below that scale, which is also the reason why the density remains close to an isothermal profile.

The radial velocity of the gas reflects the thermal properties; in general the velocities are negative due to inflow, and they are larger for $J_{21}=10^5$ compared to $J_{21}=10^3$. As we follow here the collapse of the first density peak in the halo, this behaviour makes sense as the radial infall should then approximately follow the sound speed, which is more moderate for $J_{21}=10^3$ where the temperature is lower. 
Within the central region of the halo the radial velocity then approaches zero as the very central part of the cloud is not collapsing yet. 

The vorticity structure shows two very interesting trends in this halo; first, the vorticity tends to be larger in halos with $J_{21}=10^5$ reflecting the higher velocities. In addition, for both values of $J_{21}$, we also see that the vorticity is lower for the simulation with our strongest initial magnetic field seed. 

For the sonic Mach number, we note that overall the motion is supersonic and it is the same on large scales, where also the thermal evolution is the same. On smaller scales the Mach number is larger for the runs with $J_{21}=10^3$ where the temperature is reduced, reflecting the lower sound speed. While we previously noted that also the radial velocity is reduced in the lower-temperature runs, we see here that it is not fully proportional to the difference in the sound speed. Indeed, in a free-fall collapse, the radial velocity would be the same in all simulations. While the collapse speed can be affected by thermal pressure, runs with lower temperatures will then typically show higher Mach numbers.

The evolution of the magnetic properties for the same simulations are shown in Fig.~\ref{fig:structhalo3}. Concerning the radial profile of the magnetic field strength itself, we note that in general, independent of the initial magnetic field seed, the magnetic field strength is higher in the simulations with $J_{21}=10^5$ compared to $J_{21}=10^3$. This change is consistent with the behaviour of the density, which is also higher in the case of higher $J_{21}$ and indicates that a higher degree of compression has occurred in these simulations. In addition, we also noticed before the higher vorticity in the simulations with higher $J_{21}$ as well as the smaller Mach number, which both are expected to favor a more efficient amplification of the magnetic field. 

To understand how much of the magnetic field amplification is due to compression versus dynamo amplification, we can make some simple considerations. In case of spherical symmetry, the density is expected to increase in a contracting sphere as $\rho\propto r^{-3}$. On the other hand, if we assume flux freezing, the magnetic field strength should increase as $B\propto r^{-2}$, implying that $B\propto\rho^{2/3}$ as a result of compression. In Fig.~\ref{fig:structhalo3}, we therefore plot $B/\rho^{2/3}$ in order to determine if an additional enhancement is present as a result of turbulence. We normalize this value by $B_0/\rho_{0}^{2/3}$ where $B_0$ is the initial magnetic field strength and $\rho_{0}$ is the cosmic mean baryon density at $z=100$. By plotting this quantity we can notice that on large scales this ratio is comparable independent of the value of $J_{21}$ and $B_0$, however it starts to change towards the centre. For this fixed high Jeans resolution the normalized $B/\rho^{2/3}$ is larger for simulations with stronger radiation background compared to the ones with weaker radiation background, a behaviour that is also reflected in the magnetic field strength. In addition, the lowest values of $B/\rho^{2/3}$ are obtained for the simulations with $B_0=10^{-8}$~G. \citet{Sur2012} found that the dynamo saturation in a self-gravitating system can be identified from a change in the slope of $B/\rho^{2/3}$. For the strongest seed field, we find that this ratio decreases towards the centre of the halo suggesting that the magnetic field is already close to the saturation state.

Finally, in Fig.~\ref{fig:structhalo3}, we also show the ratio of magnetic over kinetic energy for halo 1. This ratio indicates that when magnetic energy reaches a fraction of equipartition, saturation occurs. Additionally, it is strongly dependant on the Mach number and the turbulence injection mechanism giving lower values for supersonic flows with compressible turbulence \citep{brandenburg,Federrath2011,Schober2015}. For this halo, independent of the strength of the initial magnetic field seed, we find the highest ratio for higher values of $J_{21}$, the same behaviour that is present in the other magnetic quantities. For the simulations with $B_0=10^{-14}$~G (proper) it can be seen that even though the energy ratio increases towards the centre, indicating that the magnetic energy is growing at these radii, the value that it reaches is in the range of $10^{-14} - 10^{-12}$ depending on the value of $J_{21}$, therefore the motion of the halo is entirely driven by kinetic dynamics. For $B_0=10^{-10}$~G (proper) the ratio reaches values of the order of $10^{-5}-10^{-4}$, thus the magnetic field still has practically no dynamical impact. On the other hand, for our strongest initial magnetic field seed, this ratio reaches values greater than $10^{-3}$, which means that here the magnetic field starts to play a bigger role for the motion of the gas and therefore it starts to become dynamically relevant. Furthermore, we even observe that the energy ratio can reach values of 0.1 at lower radii even though not all the kinetic energy contributes to the dynamo amplification, indicating that the magnetic field is almost saturated. 

While the previous analysis has focused on the magnetic field amplification and saturation driven on the smallest scales with a fixed Jeans resolution, we also aim to study the Jeans resolution effect on the amplification and saturation of the magnetic field. For this purpose in Fig.~\ref{fig:bhalo3} we show the magnetic properties of the same simulations but varying the Jeans resolution. We note here that for different Jeans resolutions the degree of turbulence will change and consequently the physical viscosity and resistivity will not remain constant. Moreover \citet{Grete2023} have shown that even in high-resolution magnetized turbulence simulations, the MHD turbulence does not converge from an energy dynamics point of view. Therefore, we do not expect a convergent behaviour in the growth rate of the dynamo.

We start analysing column 1 for $B_0=10^{-14}$~G. Similar to what we discussed before, the magnetic field strength radial profile shows that it is higher in the atomic cooling runs ($J_{21}=10^5$), especially for our highest Jeans resolution simulations. In addition, we also noticed that the magnetic field strength for fixed value of $J_{21}$ is higher by increasing the Jeans resolution. Indeed, we found that the vorticity increases with increasing resolution per Jeans length, an expected behaviour as vorticity is known to increase on smaller scales within a turbulent cascade, providing indirect confirmation that we start to resolve the turbulence better as well as showing the potential to amplify the magnetic field through solenoidal motions (see Fig.~\ref{fig:structhalo3-2} for the physical properties of this halo for different Jeans resolutions and $B_0=10^{-14}$~G).

By plotting $B/\rho^{2/3}$ we observe again that this quantity is comparable for runs with the same Jeans resolution independent of the value of $J_{21}$ for almost all scales, however, this occurs especially for our lower Jeans resolution runs. For the higher resolution runs, particularly in this halo (for comparison you can see Fig. ~\ref{fig:structhalo2} for halo 2 and Fig.~\ref{fig:structhalo1} for halo 3), we noticed that $B/\rho^{2/3}$ is larger for the simulations with $J_{21}=10^5$ compared to the ones with $J_{21}=10^3$ considering the same Jeans resolution. Another thing that became apparent from plotting this quantity is that for fixed values of $J_{21}$, the highest values of $B/\rho^{2/3}$ are obtained for the simulations with the highest resolution per Jeans length where the magnetic field is amplified between 1 and 2 orders of magnitude above the maximum amplification expected from flux freezing.

Finally, we plot the magnetic to kinetic energy ratio, finding the highest ratio for the largest resolutions per Jeans length. However, the ratio nonetheless is $\leq 10^{-9}$ and therefore the motion of the halo is entirely driven by kinetic dynamics.

The radial profile of the magnetic properties for a simulation where we resimulate halo 1 with $B_0=10^{-10}$~G is shown in column 2 of Fig.~\ref{fig:bhalo3} for different resolutions per Jeans length and for $J_{21}=10^3$ and $J_{21}=10^5$. In the atomic cooling runs, the behaviour is similar as we found before, i.e. the largest magnetic field strengths and the largest ratio of $B/\rho^{2/3}$ is found for the highest resolution per Jeans length. On the other hand, for the simulations where molecular hydrogen cooling becomes relevant ($J_{21}=10^3$), a clear resolution dependence can no longer be seen, which may suggest that perhaps the dynamo is already saturated in those simulations. Considering that the thermal and kinetic energy densities are larger in simulations with atomic hydrogen cooling compared to molecular hydrogen cooling, it would not be unexpected if the magnetic field strength for saturation would be larger as well in the atomic cooling regime. 

We finally compare with a set of simulations of the same halo but an even higher initial magnetic field strength of $B_0=10^{-8}$~G. The radial profiles of the main magnetic quantities of halo 1 are given in column 3 from Fig.~\ref{fig:bhalo3}. In the molecular hydrogen regime, the situation remains as before in that there is no clear dependence on resolution for $B$ or $B/\rho^{2/3}$. In the atomic cooling runs, for our highest Jeans resolution simulation we note that $B/\rho^{2/3}$ decreases towards the centre. This is due to the decrease in temperature that this halo shows from a radius of about $10^{-5}$~kpc towards the interior, where molecular hydrogen starts forming more efficiently and so the cooling becomes more efficient when using 256 cells per Jeans length. Even though this happens when using the highest strength for the radiation background, this effect is not surprising and it is known from previous studies that the transition from atomic to molecular cooling does not correspond to one fixed value of $J_{21}$, but varies from halo to halo and also changing the resolution may change the properties within one specific halo and affect this transition to some degree. In addition, this decrease of the temperature leads to an increase in the sonic Mach number disfavouring the efficiency of the magnetic field amplification (see Fig.~\ref{fig:structhalo3sat2} to observe the physical properties of these runs). For the simulations with different Jeans resolution in the atomic hydrogen regime, while in a significant part of the radial profile the 128 cells per Jeans length run has the strongest $B/\rho^{2/3}$ ratio, it is comparable on some scales with the 64 cells per Jeans length simulation where it becomes higher in some radii. In the central region, the $B/\rho^{2/3}$ ratio for the 128 cells per Jeans length simulation starts to decrease while it increases in our lowest Jeans resolution simulation reaching the same magnetic field strength. Moreover, we see the change in the slope of $B/\rho^{2/3}$ and in the slope of the magnetic to kinetic energy ratio in almost all the simulations here. Even in some cases the energy ratio reaches $E_{B}/E_{K}\sim 0.1$, indicating that the magnetic field is almost saturated. It is important to note that although the dynamo-generated magnetic field may reach saturation during gravitational collapse, the magnetic field can still be amplified by gravitational compression \citep{Sur2012}.


\subsection{UV background effect}\label{UV}

In the previous subsection, we have explored the amplification of magnetic fields considering simulations with two fixed values of $J_{21}$ and varying the resolution and initial magnetic field strength to determine the magnetic field amplification and saturation coming from small scales. In this section we rather explore the dependence on $J_{21}$ itself, comparing simulations that were pursued with $128$ cells per Jeans length and an initial magnetic field strength of $B_0=10^{-14}$~G.

We start our analysis with halo 1 in the first column of Fig.~\ref{fig:bhalo3UV}, comparing the radial profiles of the magnetic properties in simulations with four different values of $J_{21}$ for a resolution of 128 cells per Jeans length. We find an enhancement of the magnetic field strength in runs with atomic hydrogen cooling on scales from $10^{-4}$~kpc $-$ $10^{-7}$~kpc. At a radius of about $10^{-2}$~kpc, the simulation with $J_{21}=100$ shows a small enhancement of the magnetic field which occurs at the same scale where a density bump is present (for the physical properties of this halo see Fig~\ref{fig:halo3compJ21}). In the ratio of $B/\rho^{2/3}$, we see a similar behaviour, on larger scales $B/\rho^{2/3}$ is larger particularly for the simulation with $J_{21}=100$, however, on smaller scales it seems to be larger for runs in the atomic hydrogen cooling regime. For the ratio of magnetic to kinetic energy we see that the lower ratio is for $J_{21}=10^3$ which is the simulation that shows the lowest magnetic field amplification and therefore the smallest magnetic energy. On the other hand, the atomic cooling runs show a lower ratio in comparison with the  $J_{21}=100$ run, however, at a scale of about $10^{-6}$~kpc the ratio of the simulation with $J_{21}=10^4$ starts to increase. This happens at the same scale where an enhancement on $B/\rho^{2/3}$ occurs.

A systematic comparison of the magnetic properties for halo~2 is shown in the second column of Fig.~\ref{fig:bhalo3UV}. In this case, we find that the simulation with $J_{21}=100$ reaches a magnetic field strength of about $10^{-6}$~G in the centre of the halo similar to what we found in halo 1. However, particularly in this halo, the other runs with a higher value of $J_{21}$ reach stronger magnetic field strengths, where unlike halo 1, the magnetic field strength reached in the simulation with $J_{21}=10^{3}$ is almost the same as in the simulation with $J_{21}=10^4$. For the magnetic field amplification $B/\rho^{2/3}$, we see the smallest $B/\rho^{2/3}$ for the simulation with $J_{21}=100$ and it keeps this lower value even on smaller scales. On scales above $10^{-6}$~kpc the $B/\rho^{2/3}$ ratio is comparable for runs with $J_{21}=10^3$ and $10^5$ while the amplification for the simulation with $J_{21}=10^4$ is smaller, however, at scales below $10^{-6}$~kpc this changes and the $J_{21}=10^4$ starts to increase reaching a magnetic field amplification similar to the simulation with $J_{21}=10^3$. In addition, we see that this halo presents the highest amplification by the small-scale dynamo without considering the case using $J_{21}=100$. The small-scale dynamo provides an additional magnetic field amplification of more than one order of magnitude above the regular amplification driven by compression. The ratio of magnetic to kinetic energy exhibits a similar behaviour where the lower ratio is for $J_{21}=100$ and the highest ratios are for the intermediate values of $J_{21}$ with a maximum value of about $10^{-9}$, two orders of magnitude higher than the energy ratios reached on halo 1.

Finally, for comparison, we present in the third column of Fig.~\ref{fig:bhalo3UV} the same quantities but for halo 3. Similar to what we found for halo 1, we find an enhancement of the magnetic field strength on scales from $10^{-4}$~kpc $-$ $10^{-7}$~kpc for the runs where the atomic hydrogen cooling dominates. On larger scales the situation is somewhat less clear and in particular the simulation with $J_{21}=100$ shows a relevant but temporary enhancement of the magnetic field. In addition, we can note that the magnetic field strength becomes comparable towards the centre independent of the value of $J_{21}$ reaching a value of about $10^{-5}$~G. In the ratio of $B/\rho^{2/3}$, we see a similar behaviour, though somewhat more pronounced for $J_{21}=100$ at a radius of about $10^{-2}$~kpc. The ratio of the magnetic to kinetic energy appears slightly lower in the $J_{21}>100$ runs.

From the comparison of these three halos, we notice that the effect of $J_{21}$ on the amplification of the magnetic field varies from halo to halo. In addition, we have to keep in mind that the amplification of the magnetic field $B/\rho^{2/3}$ can vary statistically by two orders of magnitude for virtually identical halos as shown in \cite{grete2019}, however, independently of the value of $J_{21}$ the magnetic field is efficiently amplified by many orders of magnitude (between 7 to 10).


\section{Summary and conclusions}\label{summary}

We have presented here a suite of numerical simulations modelling three different halos with masses of the order $10^7$~M$_\odot$ collapsing at redshifts $z\gtrsim12$, for which we varied the initial magnetic field strength, the resolution per Jeans length and the strength of the Lyman-Werner background parameterized via $J_{21}$. Our main goal was to study the evolution of the magnetic field and its amplification and saturation via the small-scale dynamo (on top of gravitational compression) under the different conditions explored here.

Similar to previous studies, we found that weak initial magnetic fields can be efficiently amplified via the small-scale dynamo for sufficiently high resolutions per Jeans length. Particularly we found that the strongest magnetic fields are usually obtained for the simulations with $128$ and $256$ cells per Jeans length, which were the highest resolution runs we could usually pursue. This result has been found for all three halos and independent of the value of $J_{21}$ that was employed, so it is true in both, in halos where the cooling is dominated by atomic hydrogen as well as in halos where the molecular hydrogen cooling becomes important during the evolution.

The behaviour of the magnetic field was not only analyzed via the magnetic field strength $B$ itself, but we also considered the quantity $B/ \rho^{2/3}$, as under the assumption of flux freezing and for spherically symmetric compression, we should have $B\propto\rho^{2/3}$ in the absence of further amplification mechanisms \citep[see also][]{Sur2010, Schleicher2010}. This quantity was also found to increase with increasing resolution per Jeans length when plotted as a function of radius. And very similar trends have been found in the ratio of magnetic to kinetic energy density.  

To understand when saturation occurs, we varied the magnetic field strength and found tentative evidence that saturation occurs for somewhat lower initial field strength of $10^{-10}$~G in halos where cooling is driven via molecular hydrogen, while halos cooling via atomic hydrogen may require larger initial field strength of $\sim10^{-8}$~G for saturation. This behaviour is not fully unexpected, as the radial velocities as well as the vorticity are found to be enhanced under conditions of atomic hydrogen cooling, while the respective components are smaller under molecular cooling conditions. Assuming that saturation occurs at a constant ratio of magnetic over turbulent energy, it is thus natural to expect that stronger magnetic fields will be produced in the atomic cooling regime. In fact, as our simulations show that the Mach numbers are reduced in the atomic cooling regime, it is even conceivable that also the energy ratio at which saturation occurs will be somewhat higher in the atomic cooling regime \citep[see also][]{Federrath2011}.

Our study thus confirms that magnetic field amplification via the small-scale dynamo should occur under a large range of conditions, both in the atomic and molecular hydrogen cooling regime. We also found that the saturation levels of the magnetic field are likely different for the two cases. In the future, it will be important to study also the subsequent evolution, which may include the formation of a disk where also an $\alpha-\Omega$ dynamo may operate \citep[see][]{sharda1,sharda2}. Similarly, \citet{latif8} have shown that at the transition point towards an adiabatic core, magnetic field amplification can be strongly enhanced due to the presence of shocks, and the resulting magnetic fields may reduce subsequent fragmentation. It will thus be important to further study the effects of the presence of such magnetic fields during the formation of massive objects in these halos.

\begin{acknowledgements}
      VBD acknowledges financial support from ANID (ANID-PFCHA/DOCTORADO DAAD-BECAS CHILE/62200025) as well as financial support from DAAD (DAAD/Becas Chile funding program ID 57559515). This research was supported by the North-German Supercomputing Alliance (HLRN-IV) under project grant hhp00057 and by the supercomputing infrastructure of the NLHPC (ECM-02).  DRGS gratefully acknowledges support by the ANID BASAL projects ACE210002 and FB210003, via the Millenium Nucleus NCN19-058 (TITANs) and via Fondecyt Regular (project code 1201280). DRGS also thanks for funding via the Alexander von Humboldt - Foundation, Bonn, Germany. MAL thanks the UAEU for funding via UPAR grants No. 31S390 and 12S111. RB acknowledges support by the Deutsche Forschungsgemeinschaft (DFG, German Research Foundation) under Germany’s Excellence Strategy – EXC 2121 „Quantum Universe“ – 390833306. This project has received funding from the European Union's Horizon 2020 research and innovation program under the Marie Skłodowska-Curie grant agreement No. 101030214. The visualization and analysis of this research was done thanks to the YT project, an open-source, community-developed python package for astrophysical data \citep{YTproject}.
\end{acknowledgements}

%
%

\bibliographystyle{aa} 
\bibliography{aanda} 


\begin{appendix}
    \section{Statistical comparison}\label{statistic}

   \begin{figure*}
   \centering
   \includegraphics[width=0.95\textwidth]{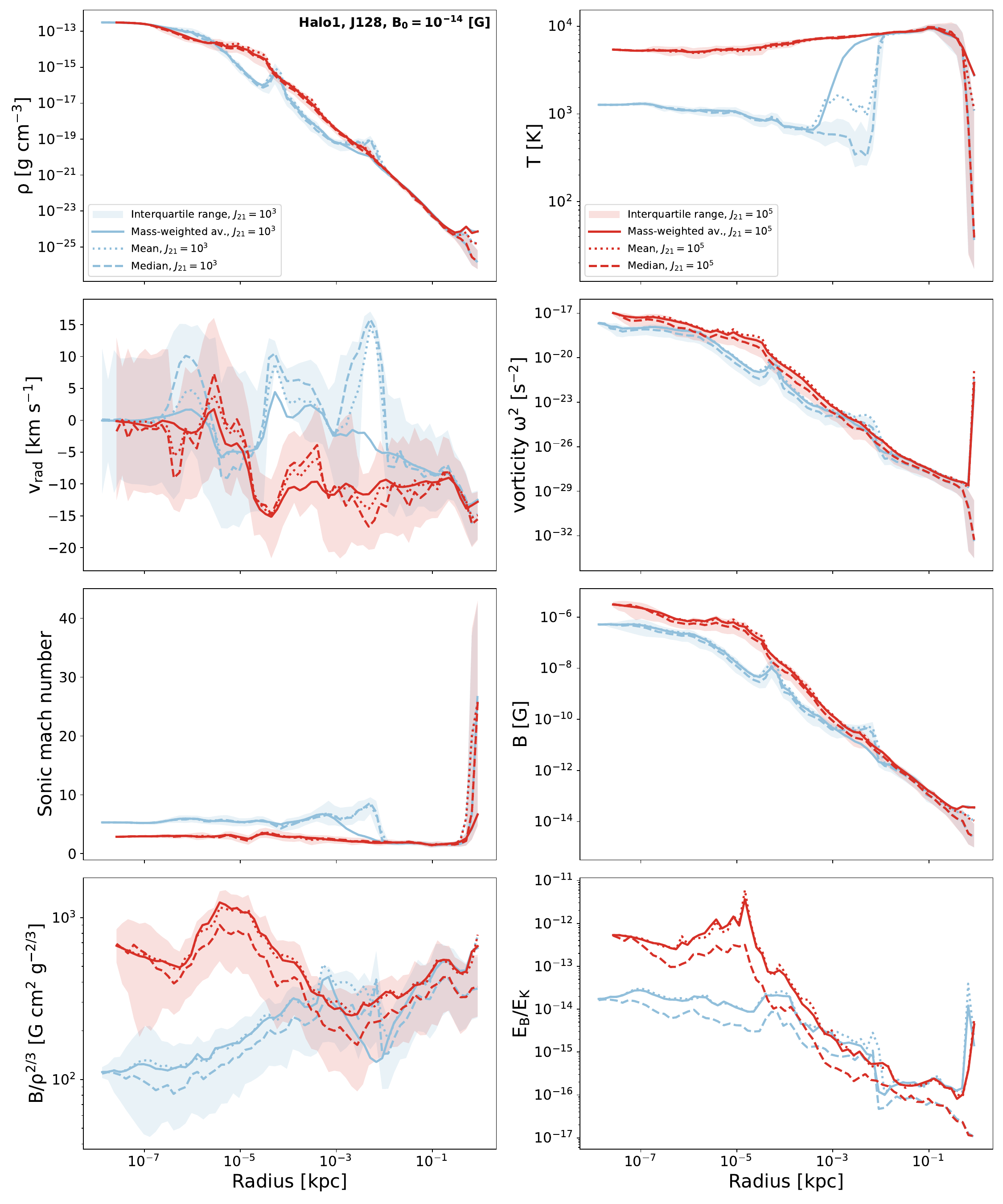}
    \caption{Spherically binned radial profiles of density, temperature, radial velocity, vorticity squared, sonic Mach number, magnetic field strength, magnetic field amplification $B/\rho^{2/3}$ and magnetic to kinetic energy density ratio for halo 1  when reaching a peak density of $3\times10^{-13}$ $g/cm^{3}$ using 128 cells per Jeans length, $J_{21}=10^5$ (red line) and $10^3$ (light blue line) and $B_0=10^{-14}\ G$ (proper). The solid lines show the mass-weighted average of the quantities, the dotted lines show the mean values, the dashed lines show the median values and the shaded areas correspond to the interquartile range of the simulations.}
    \label{fig:statistichalo1}
    \end{figure*}

    To explore how our results change with different statistical methods we plot our results for 2 different simulations. Fig.~\ref{fig:statistichalo1} show the mass-weighted average, mean, median and interquartile range of the physical and magnetic properties of halo 1 for a simulation with $J_{21}=10^3$ and $10^5$ using a Jeans resolution of 128 cells per Jeans length and an initial magnetic field of $B_0=10^{-14}~G$. It can be seen that the physical properties of the halo follows a similar behaviour when using the mass-weighted average, mean and median converging in the centre at the same point, additionally, the quantities are within the interquartile range in almost the entire radius. A difference in the values, specially in the temperature, can be seen at a scale between $10^{-4} - 10^{-2}$~kpc for the simulation in the molecular hydrogen cooling regime, where the mass-weighted average presents a smaller decrease compared to the mean and median values, however this is not surprising as it occurs in the range where the H$_2$ cooling and the H$_I$ cooling phase co-exists. For the magnetic properties we can see that even though the median tends to show smaller values compared to the mean and the mass-weighted average, the behaviour that all these quantities follow is similar. In this work, we used the mass-weighted quantities as they are the most relevant for a collapse problem, however, from a volume-weighted average point of view the results might show some differences.

    \section{Jeans resolution effect: additional halos}\label{Extrahalos}
    For comparison, here we show the results obtained for the two additional halos we simulated. The properties of these halos using different initial conditions are presented in Table~\ref{tab:table1}, \ref{tab:table3} and \ref{tab:table4}. We also provide here the physical properties of halo 1.
    
    \subsection{Halos with $B_0=10^{-14}$~G}\label{extrahalos1}

   \begin{figure*}
   \centering
   \includegraphics[width=\textwidth]{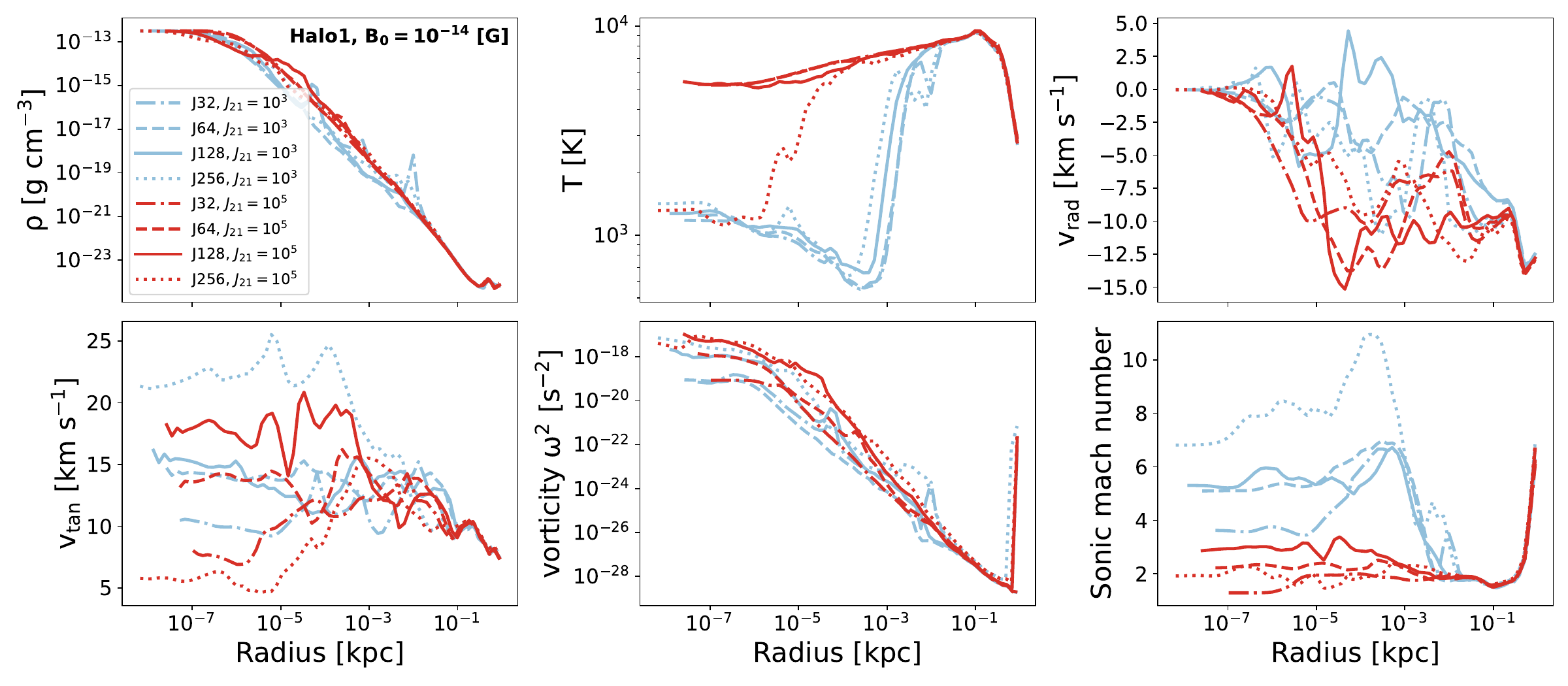}
    \caption{Mass-weighted spherically binned radial profiles of density, temperature, radial velocity, tangential velocity, vorticity squared and sonic Mach number for halo 1 when reaching a peak density of $3\times10^{-13}$ $g/cm^{3}$ using $B_0=10^{-14}\ G$ (proper). Light blue lines represent the simulations with $J_{21}=10^3$ where the cooling is driven via molecular hydrogen and red lines are for simulations with $J_{21}=10^5$ where the cooling is driven by atomic hydrogen. The different line styles represent different Jeans resolutions; dash-dotted line for 32 cells, dashed line for 64 cells, solid line for 128 cells and dotted line for 256 cells per Jeans length.}
    \label{fig:structhalo3-2}
    \end{figure*}

The radial profiles of the main physical quantities of halo 1 for different Jeans resolutions are presented in Fig.~\ref{fig:structhalo3-2}. We note that the density profile is almost independent of the resolution per Jeans length and follows an isothermal profile. The temperature for $J_{21}=10^5$ remains high, while in the runs with $J_{21}=10^3$ the gas cools down in the interior of the halo due to the molecular hydrogen. While our results mostly remain qualitatively similar, a difference can be noted in the simulation with $J_{21}=10^5$ and 256 cells per Jeans length, where the temperature initially stays high, but then starts dropping from a radius of about $10^{-4}$~kpc towards small scales. 

The radial velocity for the atomic cooling regime are negative due to inflow and larger compared to the runs in the molecular hydrogen cooling regime. We note that the resolution per Jeans length introduces a relevant scatter in the radial velocities, though without following a systematic trend. The tangential velocity shows a very similar behaviour and again we note that it is typically larger in the simulations with $J_{21}=10^5$ for essentially the same reasons, however, particularly for this halo, this is not the case in all the resolutions, specially for our highest one where the tangential velocity for $J_{21}=10^3$ is larger compared to $J_{21}=10^5$; only on large scales the tangential velocity is rather driven by effects from the cosmological environment and therefore similar in all runs. We note also here the significant scatter introduced by the resolution per Jeans length.

The vorticity is larger in halos dominated by atomic cooling and independent of the value of $J_{21}$ it increases with increasing the Jeans resolution. For the sonic Mach number, we finally note that it is larger for the runs in the molecular hydrogen regime where the temperature is reduced.

   \begin{figure*}
   \centering
   \includegraphics[width=\textwidth]{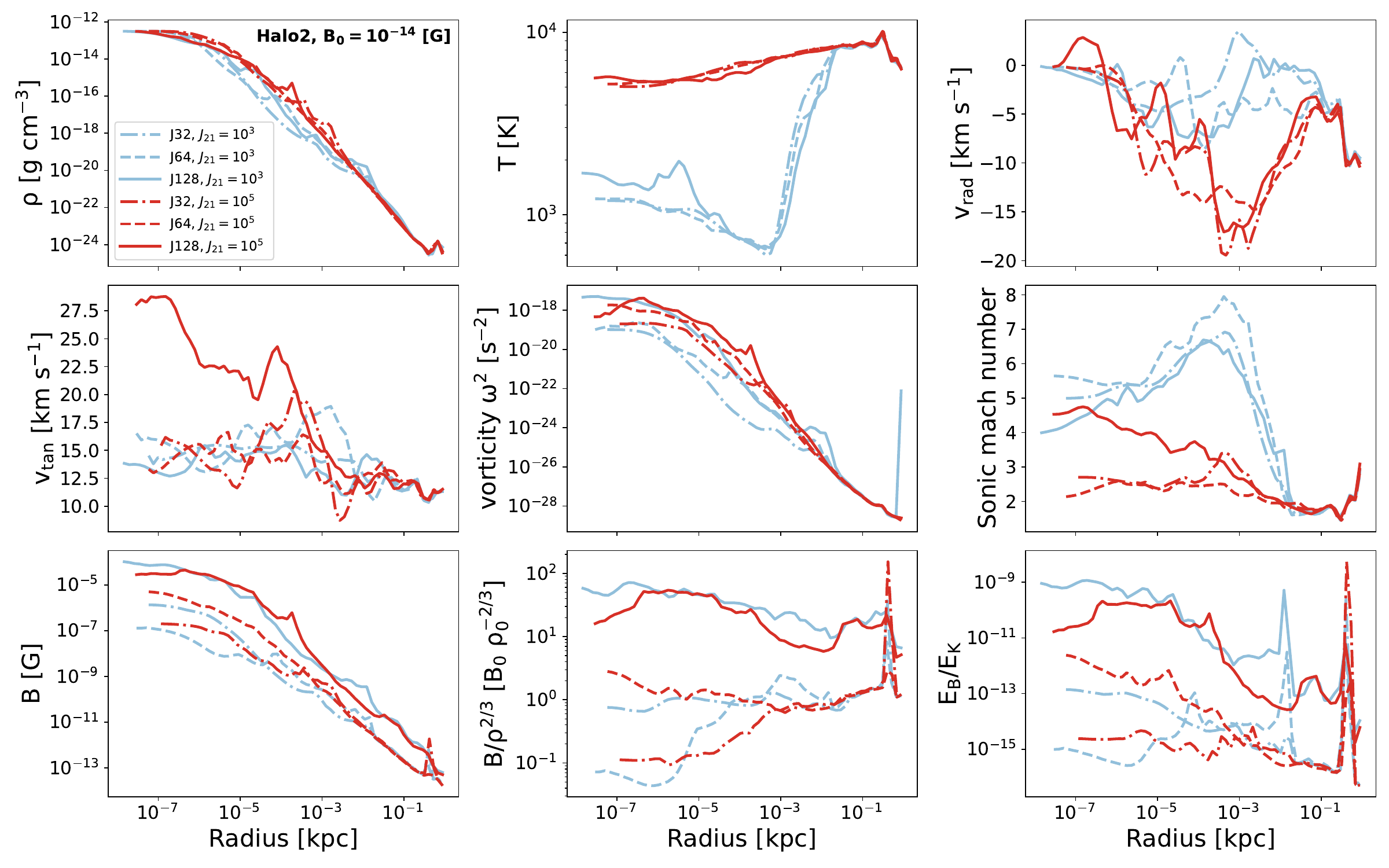}
    \caption{Mass-weighted spherically binned radial profiles of density, temperature, radial velocity, tangential velocity, vorticity squared, sonic Mach number, magnetic field strength, magnetic field amplification $B/\rho^{2/3}$ and magnetic to kinetic energy density ratio for halo 2 when reaching a peak density of $3\times10^{-13}$ $g/cm^{3}$ using $B_0=10^{-14}\ G$ (proper). Light blue lines represent the simulations with $J_{21}=10^3$ where the cooling is driven via molecular hydrogen and red lines are for simulations with $J_{21}=10^5$ where the cooling is driven by atomic hydrogen. The different line styles represent different Jeans resolutions; dash-dotted line for 32 cells, dashed line for 64 cells and solid line for 128 cells per Jeans length.}
    \label{fig:structhalo2}
    \end{figure*}

Fig.~\ref{fig:structhalo2} shows the radial profile of the main physical and magnetic quantities of halo 2. We note that the overall behaviour is very similar to what we found in halo 1, with the density approximately following an isothermal profile, and the H$_2$ cooling kicking in for cases with $J_{21}=10^3$. The behaviour of temperatures affects the radial velocity, tangential velocity, vorticity and sonic Mach number in a similar way as already noted for halo 1. As for the magnetic properties, the effect of the amplification of the magnetic field is even more pronounced for this halo and the two runs with a resolution of 128 cells per Jeans length show the larger magnetic field strength in the radial profile and also the largest ratios of $B/\rho^{2/3}$. The same is true also for the ratio of magnetic over kinetic energy.

   \begin{figure*}
   \centering
   \includegraphics[width=\textwidth]{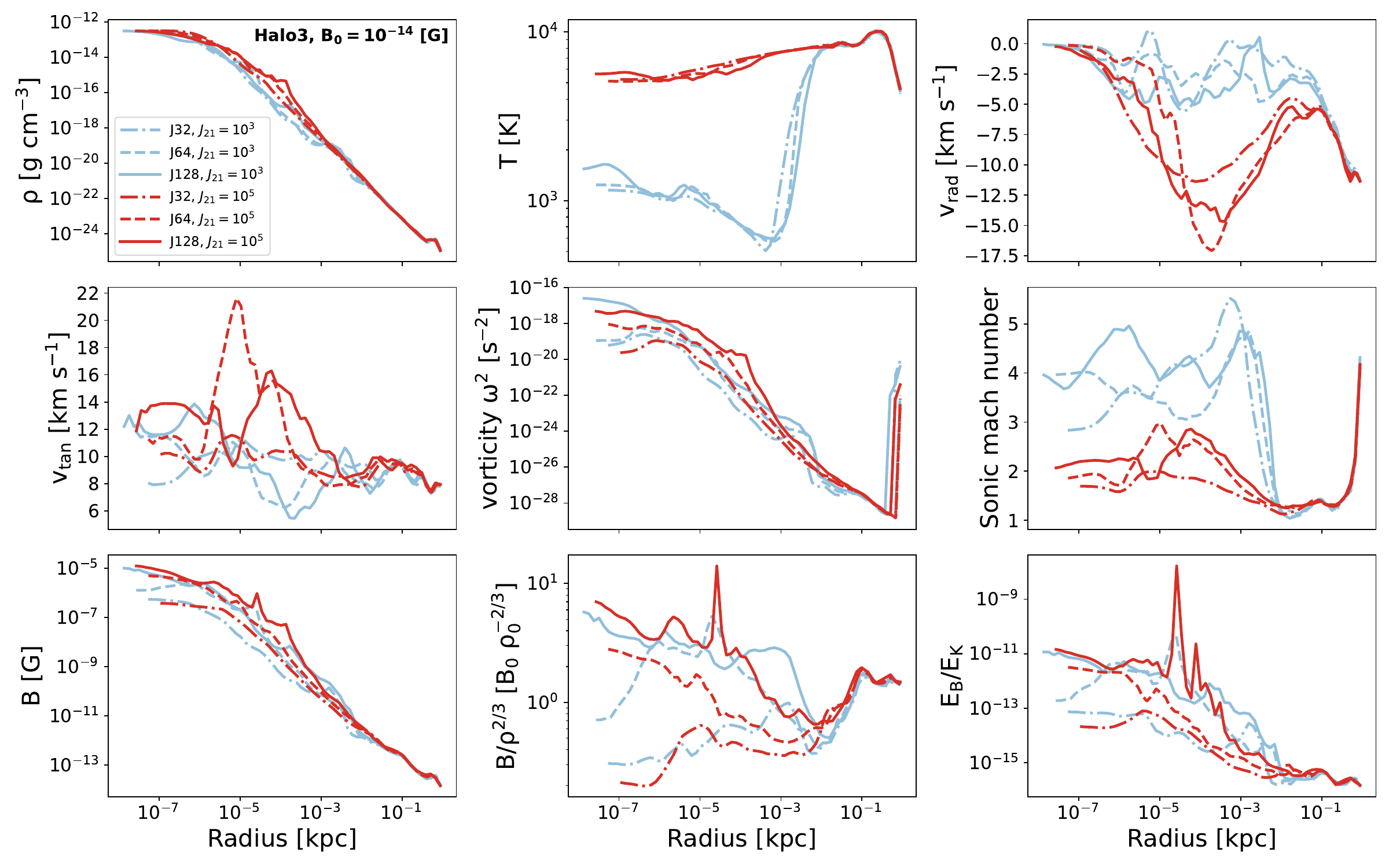}
    \caption{Mass-weighted spherically binned radial profiles of density, temperature, radial velocity, tangential velocity, vorticity squared, sonic Mach number, magnetic field strength, magnetic field amplification $B/\rho^{2/3}$ and magnetic to kinetic energy density ratio for halo 3 when reaching a peak density of $3\times10^{-13}$ $g/cm^{3}$ using $B_0=10^{-14}\ G$ (proper). Light blue lines represent the simulations with $J_{21}=10^3$ where the cooling is driven via molecular hydrogen and red lines are for simulations with $J_{21}=10^5$ where the cooling is driven by atomic hydrogen. The different line styles represent different Jeans resolutions; dash-dotted line for 32 cells, dashed line for 64 cells and solid line for 128 cells per Jeans length.}
    \label{fig:structhalo1}
    \end{figure*}
  
Quite similarly, we have analyzed the radial structure and magnetic properties of halo 3 which are shown in Fig~\ref{fig:structhalo1}. The results are very similar to what we obtained in halo 1 and 2. In this halo we can see more clearly that the infall velocity is enhanced in the simulations with higher $J_{21}$, which is also reflected in the tangential velocities and in the vorticity. The sonic mach number is larger in the simulations with lower $J_{21}$ due to the lower temperature. For the magnetic quantities we again obtain a picture consistent with the results from the other halos; particularly we found the largest magnetic field strengths in the highest resolution runs. In addition, in this halo we can better appreciate how runs with the same resolution per Jeans length show a similar behavior of $B/\rho^{2/3}$, regardless of the value of $J_{21}$.

\subsection{Halos with $B_0=10^{-10}$~G}\label{extrahalos2}

   \begin{figure*}
   \centering
   \includegraphics[width=\textwidth]{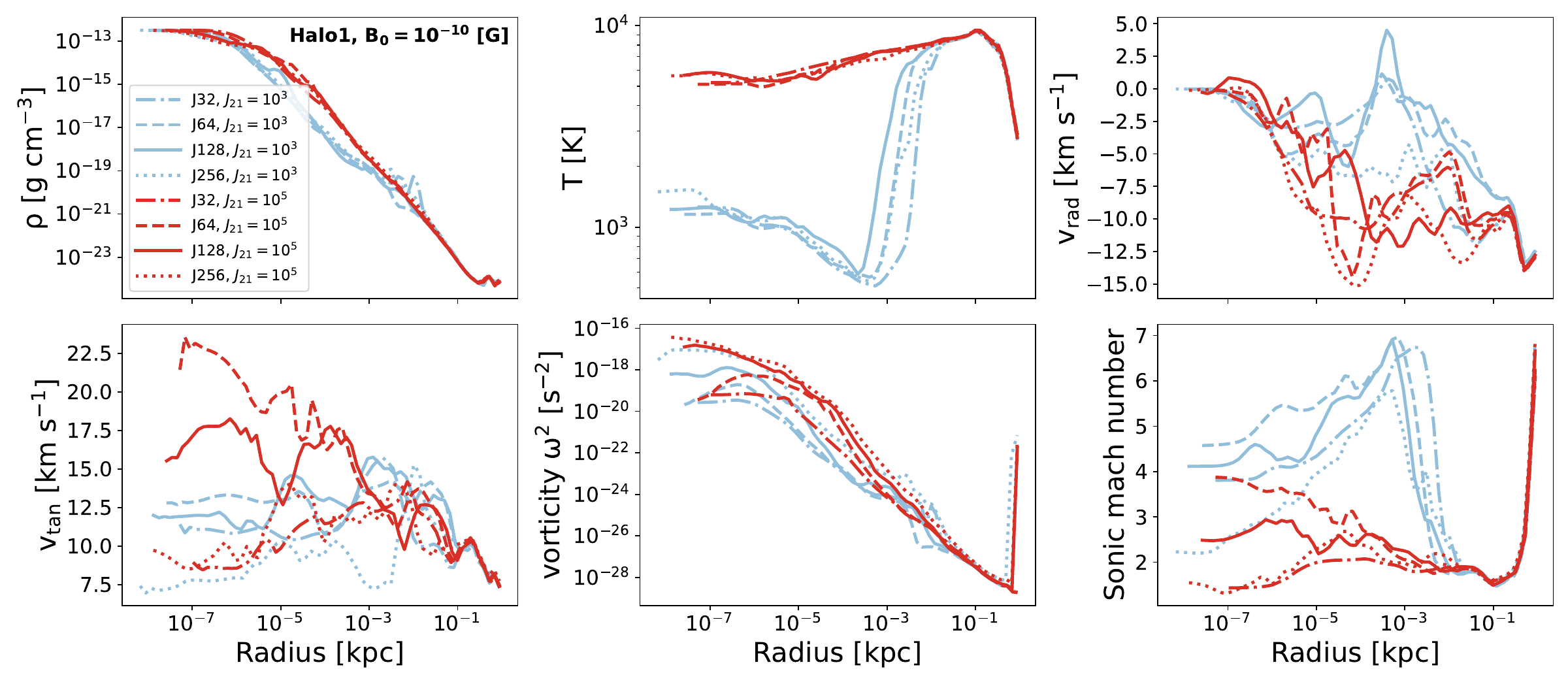}
    \caption{Mass-weighted spherically binned radial profiles of density, temperature, radial velocity, tangential velocity, vorticity squared and sonic Mach number for halo 1 when reaching a peak density of $3\times10^{-13}$ $g/cm^{3}$ using $B_0=10^{-10}\ G$ (proper). Light blue lines represent the simulations with $J_{21}=10^3$ where the cooling is driven via molecular hydrogen and red lines are for simulations with $J_{21}=10^5$ where the cooling is driven by atomic hydrogen. The different line styles represent different Jeans resolutions; dash-dotted line for 32 cells, dashed line for 64 cells, solid line for 128 cells and dotted line for 256 cells per Jeans length.}
    \label{fig:structhalo3sat}
    \end{figure*}
We present in Fig.~\ref{fig:structhalo3sat} the radial profile of the physical quantities for halo 1 which we resimulate using a stronger initial magnetic field. Within the physical properties of the halo, particularly density, temperature and the velocity components, the behaviour remains very similar to what we discussed before. However, in the simulations with $J_{21}=10^5$ and 256 cells per Jeans length, we observe a difference compared to the corresponding simulation with a weaker initial magnetic field. While the temperature shows a drop from a radius of about $10^{-4}$~kpc towards the centre when using an initial magnetic field of $B_0=10^{-14}$~G (as shown in Fig.~\ref{fig:structhalo3-2}), here we can see that the temperature remains at high values as in all simulations with $J_{21}=10^5$. In addition, we notice that the tangential velocity and the sonic Mach number are smaller in the simulation with $J_{21}=10^3$ and 256 cells per Jeans length compared to the same simulation but using $B_0=10^{-14}$~G. Potentially, this could be consider to be a result of angular momentum transport due to magnetic braking, or potentially it is part of a change in the non-linear evolution induced by the presence of a stronger magnetic field.

   \begin{figure*}
   \centering
   \includegraphics[width=\textwidth]{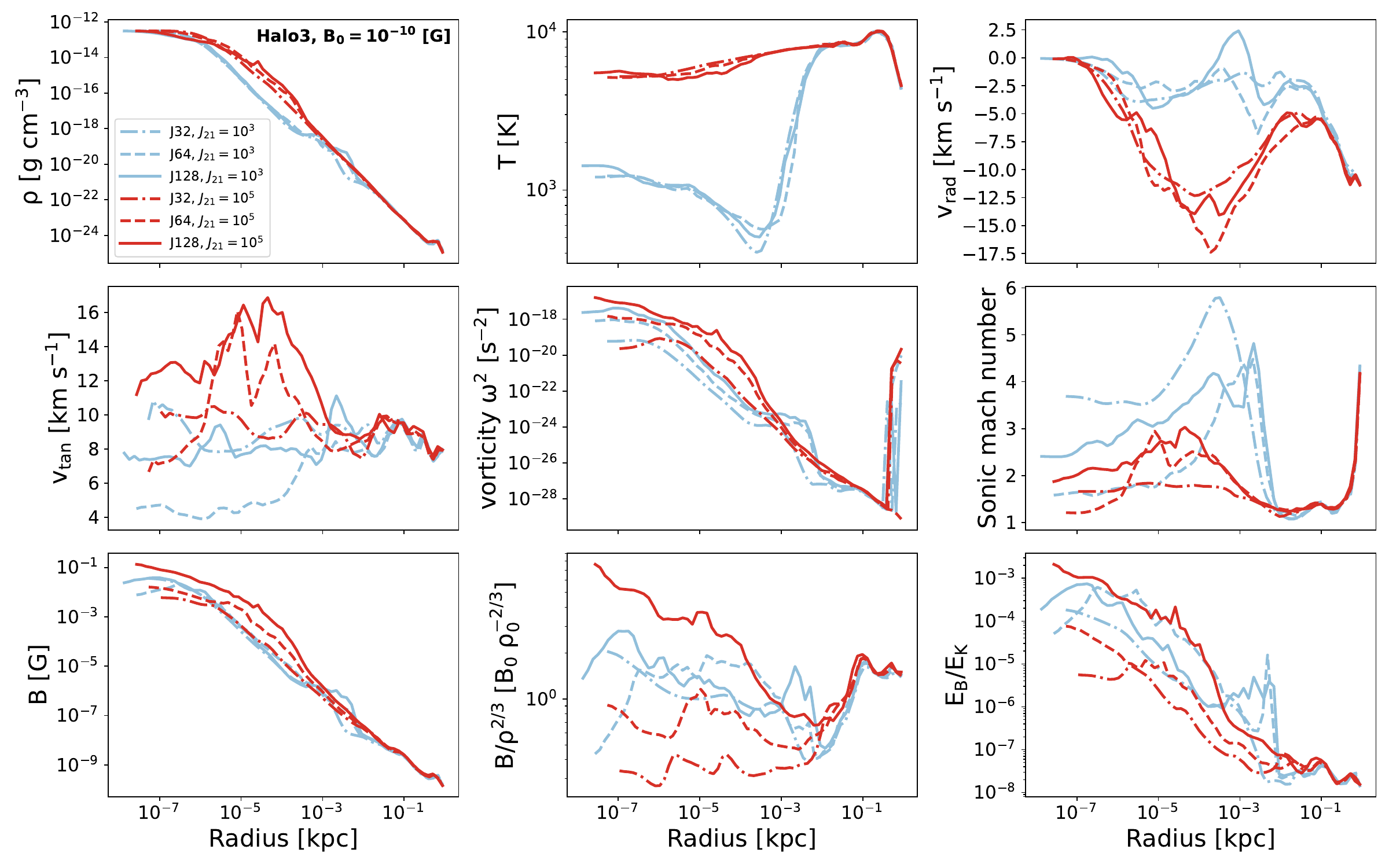}
    \caption{Mass-weighted spherically binned radial profiles of density, temperature, radial velocity, tangential velocity, vorticity squared, sonic Mach number, magnetic field strength, magnetic field amplification $B/\rho^{2/3}$ and magnetic to kinetic energy density ratio for halo 3 when reaching a peak density of $3\times10^{-13}$ $g/cm^{3}$ using $B_0=10^{-10}\ G$ (proper). Light blue lines represent the simulations with $J_{21}=10^3$ where the cooling is driven via molecular hydrogen and red lines are for simulations with $J_{21}=10^5$ where the cooling is driven by atomic hydrogen. The different line styles represent different Jeans resolutions; dash-dotted line for 32 cells, dashed line for 64 cells and solid line for 128 cells per Jeans length.}
    \label{fig:structhalo1sat}
    \end{figure*}
  The radial profiles for halo 3 using $B_0=10^{-10}$~G are shown in Fig.~\ref{fig:structhalo1sat} where the behavior of the physical properties of this halo is consistent with our findings for halo 1. The magnetic field related properties of this halo, again, show similar results compared to halo 1. For simulations in the atomic cooling regime we find the largest magnetic fields and $B/\rho^{2/3}$ ratio in the higher Jeans resolution runs, however, in the molecular hydrogen cooling regime the Jeans resolution dependence disappears which could imply that these simulations are reaching the saturation state.

\subsection{Halos with $B_0=10^{-8}$~G}\label{extrahalos3}

   \begin{figure*}
   \centering
   \includegraphics[width=\textwidth]{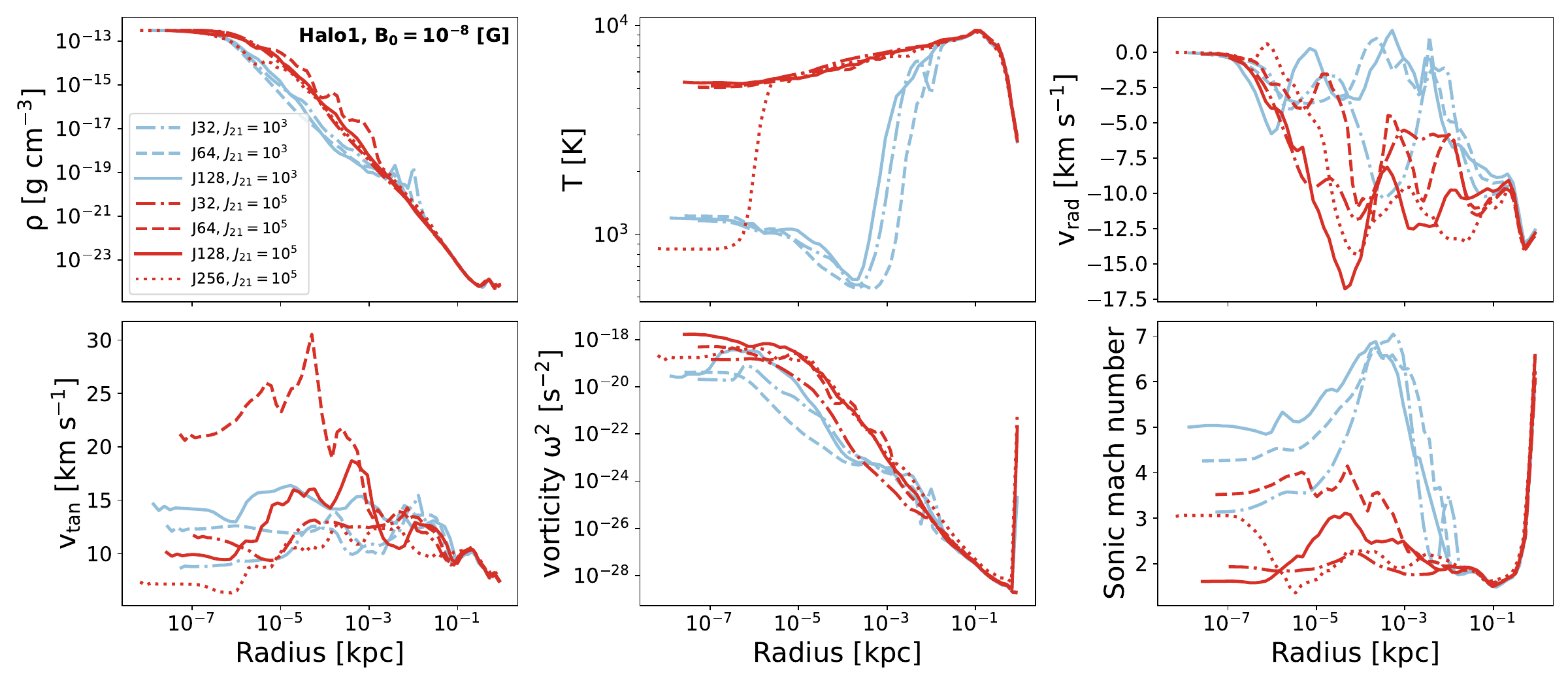}
    \caption{Mass-weighted spherically binned radial profiles of density, temperature, radial velocity, tangential velocity, vorticity squared and sonic Mach number for halo 1 when reaching a peak density of $3\times10^{-13}$ $g/cm^{3}$ using $B_0=10^{-8}\ G$ (proper). Light blue lines represent the simulations with $J_{21}=10^3$ where the cooling is driven via molecular hydrogen and red lines are for simulations with $J_{21}=10^5$ where the cooling is driven by atomic hydrogen. The different line styles represent different Jeans resolutions; dash-dotted line for 32 cells, dashed line for 64 cells, solid line for 128 cells and dotted line for 256 cells per Jeans length.}
    \label{fig:structhalo3sat2}
    \end{figure*}
We show here in Fig.~\ref{fig:structhalo3sat2} the main physical properties of the simulations of halo 1 but with an even higher initial magnetic field strength of $B_0=10^{-8}$~G. These properties again resemble the behaviour that we previously discussed, however, again a difference in temperature can be observed in the simulation with $J_{21}=10^5$ and 256 cells per Jeans length. The temperature remains high up to a radius of about $10^{-5}$~kpc where the gas cools down to smaller scales. We observe that the gas stays hot for smaller radii compared to the same simulation using $B_0=10^{-14}$~G where the temperature drops from a radius of $10^{-4}$~kpc (see Fig.~\ref{fig:structhalo3-2}). The temperature difference also leads to an increase in the sonic Mach number from a scale of $10^{-5}$~kpc, the same scale where the temperature starts to decrease in the simulation with $J_{21}=10^5$ and 256 cells per Jeans length.

   \begin{figure*}
   \centering
   \includegraphics[width=\textwidth]{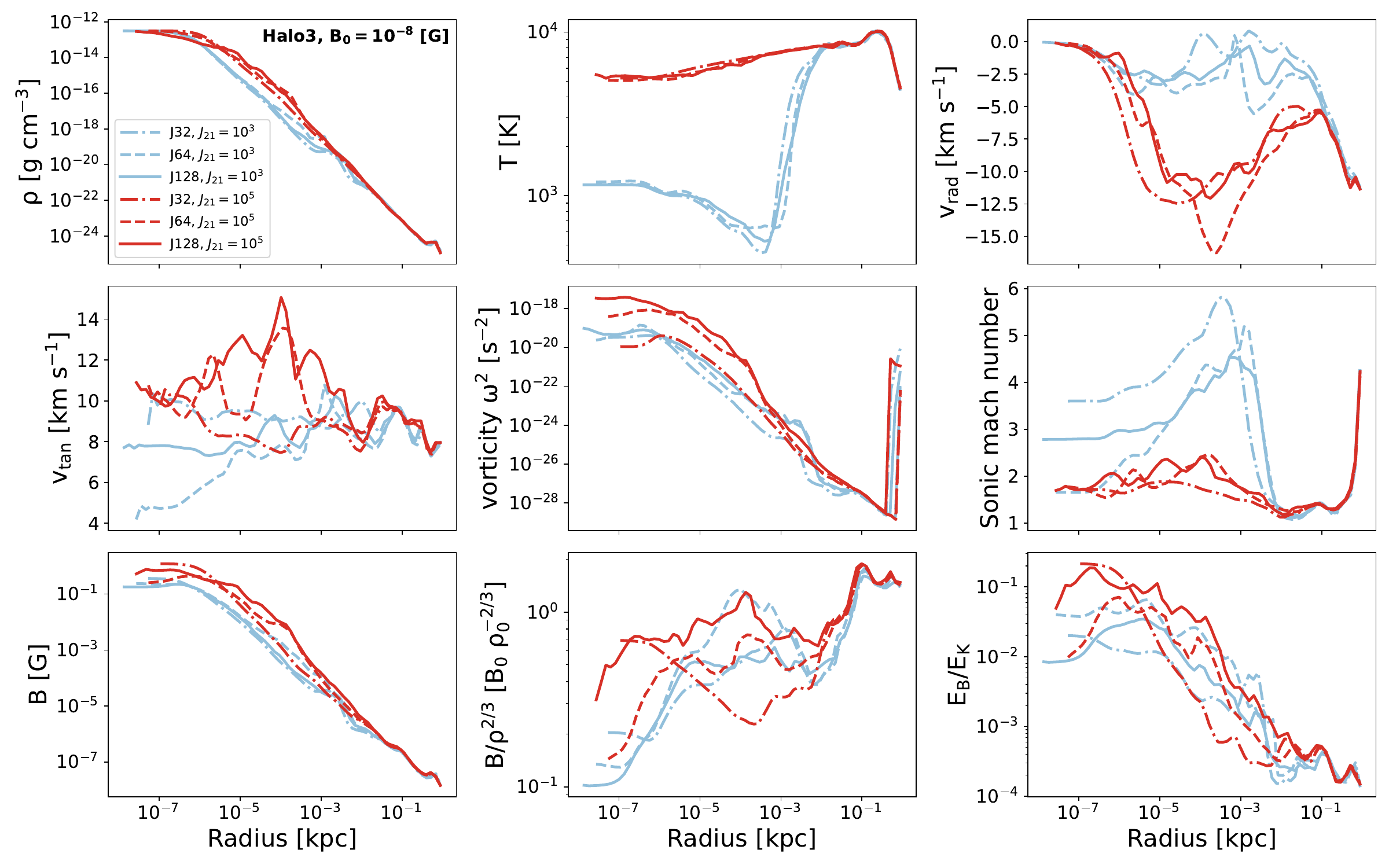}
    \caption{Mass-weighted spherically binned radial profiles of density, temperature, radial velocity, tangential velocity, vorticity squared, sonic Mach number, magnetic field strength, magnetic field amplification $B/\rho^{2/3}$ and magnetic to kinetic energy density ratio for halo 3 when reaching a peak density of $3\times10^{-13}$ $g/cm^{3}$ using $B_0=10^{-8}\ G$ (proper). Light blue lines represent the simulations with $J_{21}=10^3$ where the cooling is driven via molecular hydrogen and red lines are for simulations with $J_{21}=10^5$ where the cooling is driven by atomic hydrogen. The different line styles represent different Jeans resolutions; dash-dotted line for 32 cells, dashed line for 64 cells and solid line for 128 cells per Jeans length.}
    \label{fig:structhalo1sat2}
    \end{figure*}

  Finally, we present in Fig~\ref{fig:structhalo1sat2} the main physical and magnetic properties of halo 3 using an initial magnetic field of $B_0=10^{-8}$~G where again the behaviour is consistent with the results of the previous simulations. From the magnetic properties of the halo we can observe the no clear resolution dependence of the magnetic field strength and $B/\rho^{2/3}$ in the runs dominated by molecular hydrogen cooling remains the same. Also in the atomic hydrogen regime, while the highest resolution simulation still has the strongest magnetic field in a significant part of the radial profile, in the central region it becomes weaker compared to the lowest resolution run, similar to halo 1. This is also reflected in $B/\rho^{2/3}$ and energy ratios. Even though we cannot fully exclude that some turbulent magnetic field amplification still happens in the simulation in the atomic cooling regime, the dependency loss of the magnetic field strength and amplification on the Jeans resolution plus the fact that the magnetic to kinetic energy density ratio decreases towards the centre reaching values of about 0.1 indicates that the simulations in the atomic cooling regime may be close to the stage of saturation.

    \section{UV background effect: physical properties}\label{extrahalosUV}

   \begin{figure*}
   \centering
   \includegraphics[width=\textwidth]{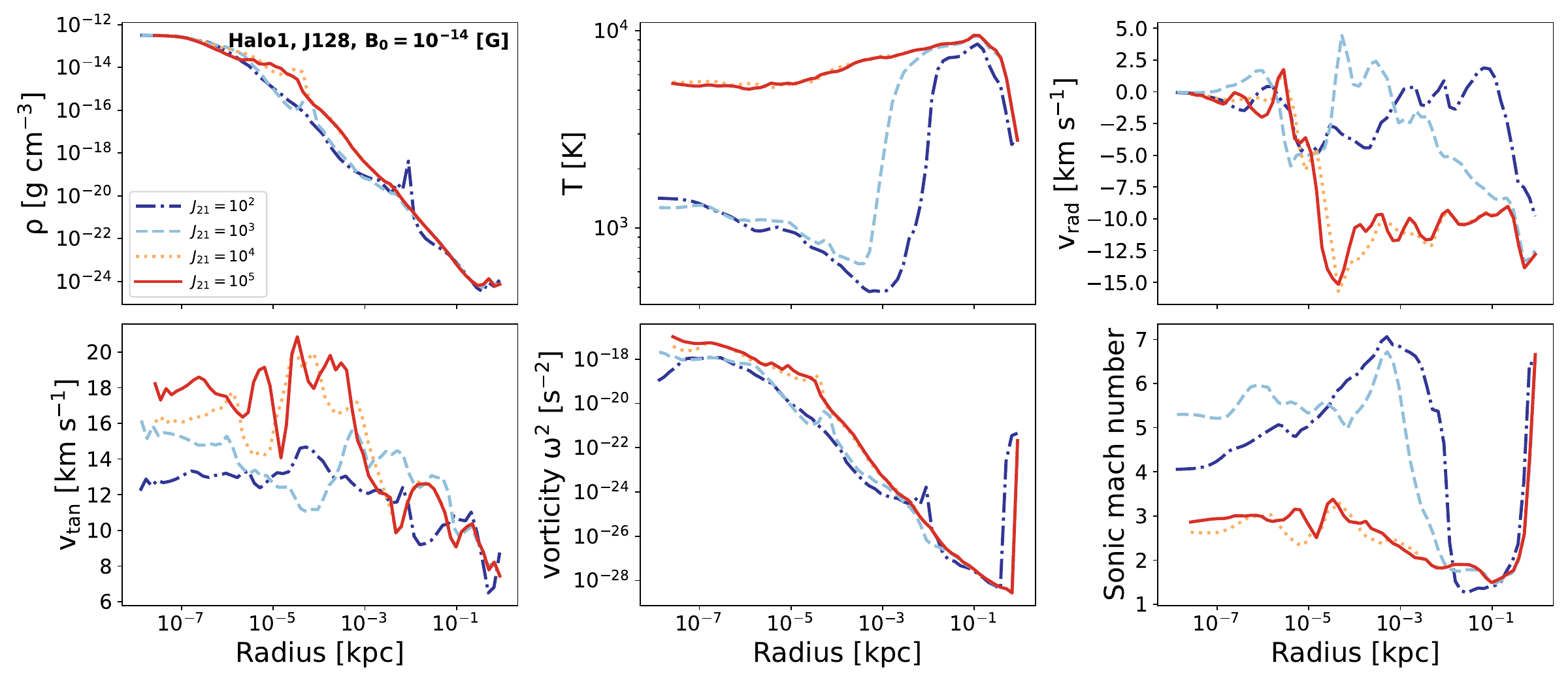}
    \caption{Mass-weighted spherically binned radial profiles of density, temperature, radial velocity, tangential velocity, vorticity squared and sonic Mach number for halo 1 when reaching a peak density of $3\times10^{-13}$ $g/cm^{3}$ using a fixed Jeans resolution of 128 cells and $B_0=10^{-14}$~[G] (proper). The dash-dotted blue line is for $J_{21}=10^2$, the dashed light blue line is for $J_{21}=10^3$, the dotted orange line is for $J_{21}=10^4$ and the solid red line is for $J_{21}=10^5$.}
    \label{fig:halo3compJ21}
    \end{figure*}

   Here we show the physical properties of the simulations presented in section~\ref{UV}, where we compare the radial profiles of our three different halos by varying the values of $J_{21}$ with a fixed Jeans resolution.
   
   The physical properties of halo 1 are shown in Fig.~\ref{fig:halo3compJ21}. The radial profile of the density distribution still approximately corresponds to an isothermal profile, with a bump due to inhomogeneities on a scale of about $\sim10^{-4}$~kpc in the simulation with $J_{21}=10^5$ and $J_{21}=10^4$. In general, on scales between $10^{-3}$~kpc and $10^{-6}$~kpc, the density tends to be higher for larger values of $J_{21}$. This can be understood as a result of the increase in temperature due to the dissociation of H$_2$, which leads to an increase of the Jeans mass and effectively means that a larger amount of mass has collapsed once a certain peak density is reached. In the temperature profile we find that H$_2$ cooling becomes possible for $J_{21}\leq10^3$, while for larger values cooling is regulated via atomic hydrogen. The profiles of radial velocity, tangential velocity, vorticity and sonic Mach number follow the previously discussed behaviour for runs with atomic hydrogen and molecular hydrogen cooling, respectively.

\begin{figure*}
   \centering
   \includegraphics[width=\textwidth]{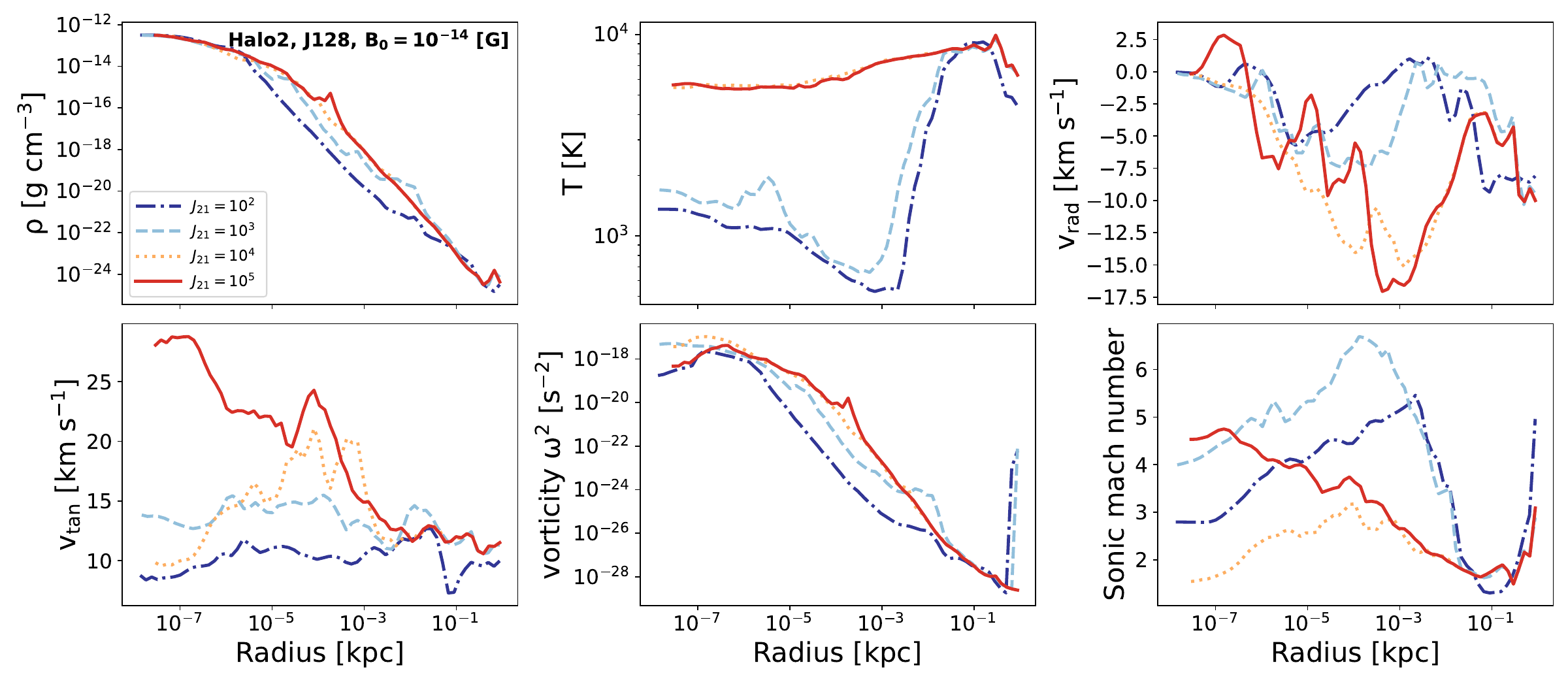}
    \caption{Mass-weighted spherically binned radial profiles of density, temperature, radial velocity, tangential velocity, vorticity squared and sonic Mach number for halo 2 when reaching a peak density of $3\times10^{-13}$ $g/cm^{3}$ using a fixed Jeans resolution of 128 cells and $B_0=10^{-14}$~[G] (proper). The dash-dotted blue line is for $J_{21}=10^2$, the dashed light blue line is for $J_{21}=10^3$, the dotted orange line is for $J_{21}=10^4$ and the solid red line is for $J_{21}=10^5$.}
    \label{fig:halo2compJ21}
    \end{figure*}

\begin{figure*}
   \centering
   \includegraphics[width=\textwidth]{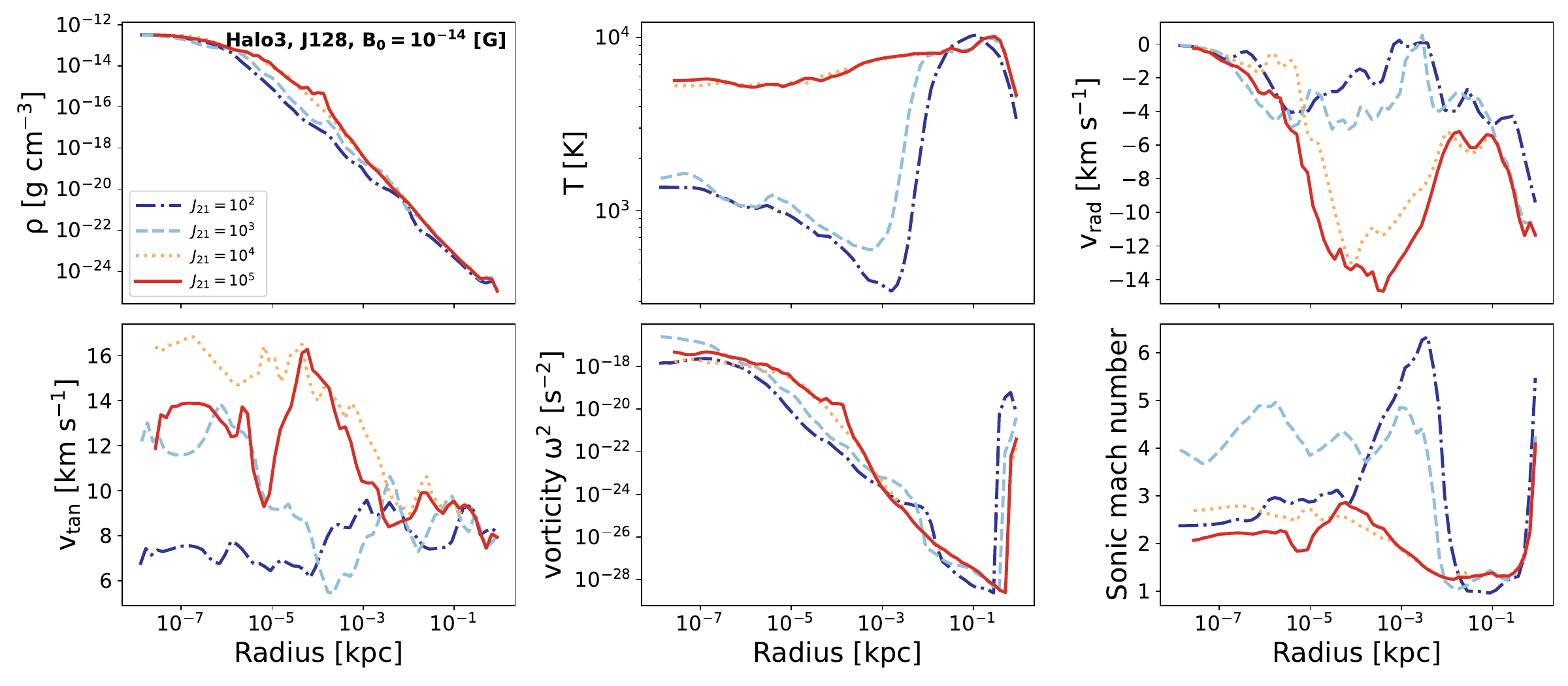}
    \caption{Mass-weighted spherically binned radial profiles of density, temperature, radial velocity, tangential velocity, vorticity squared and sonic Mach number for halo 3 when reaching a peak density of $3\times10^{-13}$ $g/cm^{3}$ using a fixed Jeans resolution of 128 cells and $B_0=10^{-14}$~[G] (proper). The dash-dotted blue line is for $J_{21}=10^2$, the dashed light blue line is for $J_{21}=10^3$, the dotted orange line is for $J_{21}=10^4$ and the solid red line is for $J_{21}=10^5$.}
    \label{fig:halo1compJ21}
    \end{figure*}

   Overall, halo 2 and halo 3 present similar trends on its physical properties as shown in Fig~\ref{fig:halo2compJ21} and Fig.~\ref{fig:halo1compJ21}, respectively. The density profile follows the approximately isothermal profile. The temperature remains high in the simulations with $J_{21}\geq10^4$ in the regime of atomic hydrogen cooling, while molecular cooling becomes active in simulations with  $J_{21}\leq10^3$ leading to lower temperatures in the interior. The infall velocity and tangential velocity are higher in the simulations in the atomic cooling regime opposite to the sonic mach number which is lower when the atomic cooling dominates the cooling. For the sonic mach number, a difference is observed for the simulation using $J_{21}=10^5$ in halo 2, where instead of maintaining a low value at small radii like halo 1 and 3, it starts to increase towards the centre (see Fig.~\ref{fig:halo2compJ21}). For the vorticity, the difference is minor, but it appears somewhat enhanced on intermediate scales ($10^{-3}$~kpc~$-$~$10^{-6}$~kpc) in the presence of atomic hydrogen cooling in both halos.

\end{appendix}

\end{document}